\documentclass[%
reprint,
superscriptaddress,
frontmatterverbose,
amsmath,amssymb,
aps,
prb,
]{revtex4-2}

\usepackage{longtable}
\usepackage{float}
\usepackage{graphicx}
\usepackage{xcolor}
\usepackage[ruled, lined, linesnumbered, commentsnumbered, longend]{algorithm2e}
\usepackage{enumerate}
\usepackage{amsmath,amssymb,amsthm,amsfonts}
\usepackage{bbm}
\usepackage{bm}
\usepackage{url}
\usepackage{braket}
\renewcommand{\vec }{\bm}
\renewcommand{\mathbf}{\bm}

\usepackage{xcolor}
\definecolor{midnightblue}{cmyk}{1,1,0,0.1}
\definecolor{forestgreen}{cmyk}{0.76,0,0.26,0.5}

\usepackage{hyperref}
\hypersetup{
    unicode=false,          
    pdftoolbar=true,        
    pdfmenubar=true,        
    pdffitwindow=false,     
    pdfstartview={FitH},    
    pdftitle={My title},    
    pdfauthor={Author},     
    pdfsubject={Subject},   
    pdfcreator={Creator},   
    pdfproducer={Producer}, 
    pdfkeywords={keyword1} {key2} {key3}, 
    pdfnewwindow=true,      
    colorlinks=true,       
    linkcolor=midnightblue,          
    citecolor=midnightblue,        
    filecolor=midnightblue,      
    urlcolor=midnightblue,          
}

\SetAlFnt{\sffamily}

\SetAlCapFnt{\normalfont\sffamily\large}

\begin{document}

\title{A Recursive Hybrid Tetrahedron Method for Brillouin-zone Integration}

\author{Kun Dong}
\affiliation{International Center for Quantum Materials, School of Physics, Peking University, Beijing 100871, China}

\author{Yihao Lin}
\affiliation{International Center for Quantum Materials, School of Physics, Peking University, Beijing 100871, China}

\author{Xiaoqiang Liu}
\affiliation{Hefei National Research Center for Physical Sciences at the Microscale, University of Science and Technology of China, Hefei 230026, China}

\author{Jiechao Feng}
\affiliation{International Center for Quantum Materials, School of Physics, Peking University, Beijing 100871, China}

\author{Ji Feng}\email{jfeng11@pku.edu.cn}
\affiliation{International Center for Quantum Materials, School of Physics, Peking University, Beijing 100871, China}
\affiliation{Hefei National Laboratory, Hefei 230088, China}

\begin{abstract}
A recursive extension of the hybrid tetrahedron method for Brillouin-zone integration is proposed, allowing
iterative tetrahedron refinement and significantly reducing the error from the linear tetrahedron method. The Brillouin-zone integral is expressed as a weighted sum on the initial grid, with integral weights collected recursively from the finest grid. Our method is capable of simultaneously handling multiple singularities in the integrand and thus may provide practical solutions to various Brillouin-zone integral tasks encountered in realistic calculations, including the computation of response and spectral function with superior sampling convergence. We demonstrate its effectiveness through numerical calculations of the density response functions of two model Hamiltonians and one real material system, the face-centered cubic cobalt.

\end{abstract}

\maketitle

\section{Introduction\label{intro}}
Brillouin-zone (BZ) integration has played an indispensable role in modern electronic structure methods for crystalline solids, especially in evaluating the thermodynamic and kinetic properties of periodic systems. In its conventional implementation, special care is required for metallic systems at $T=0\,\mathrm{K}$, where discontinuities in the integrand due to partial occupation of the BZ can lead to poor convergence if not properly handled. Two methods are widely used to address these discontinuities: the broadening method\cite{methfessel1989a} replaces zero-temperature occupancy $f_{\vec k}=\theta(\varepsilon_F-\varepsilon_{\vec k})$ by its smooth approximant, while the linear tetrahedron method\cite{jepson1971,lehmann1972} integrates only within the occupied region.
Though neat in its management of discontinuities, the linear tetrahedron method is limited by linear interpolation. It converges slowly at the 2nd order in characteristic $\vec k$-spacing and requires a substantial number of sampling points.

A few techniques have been proposed to remedy this problem. For example, using quadratic interpolation instead leads to the direct quadratic tetrahedron method\cite{wiesenekker1988,wiesenekker1991}. 
In the celebrated Bl\"ochl's ``improved tetrahedron method", the integration error due to the nonzero curvature of the integrand is corrected by a Fermi surface integral by virtue of the Stokes' theorem\cite{blochl1994}. Yet, for Bl\"ochl's method to be applicable, the integrand (apart from a factor $f_{\vec k}$) 
needs to be a regular function. This limits its application to the calculation of response and spectral functions, where an integrand of the form $A_{\vec k}/D_{\vec k}$ is involved, and the denominator $D_{\vec k}$ typically contains zeros in the BZ. Lambin and Vigneron's (linear) analytical tetrahedron method\cite{lambin1984} analytically integrates $A_{\vec k}/D_{\vec k}$ with a linearly interpolated numerator and denominator but suffers the same slow convergence as the linear tetrahedron method. More recently, an ``improved tetrahedron method" for calculating response functions was proposed, in which $A_{\vec k}$ and  $D_{\vec k}$ are replaced with so-called leveled linear approximants, constructed via fitting their third-order interpolants\cite{kawamura2014}.
However, there is no theoretically sound justification for using the leveled denominator, which apparently has different poles than the true integrand.

\begin{figure}[hb]
    \centering
    \includegraphics[width=8cm]{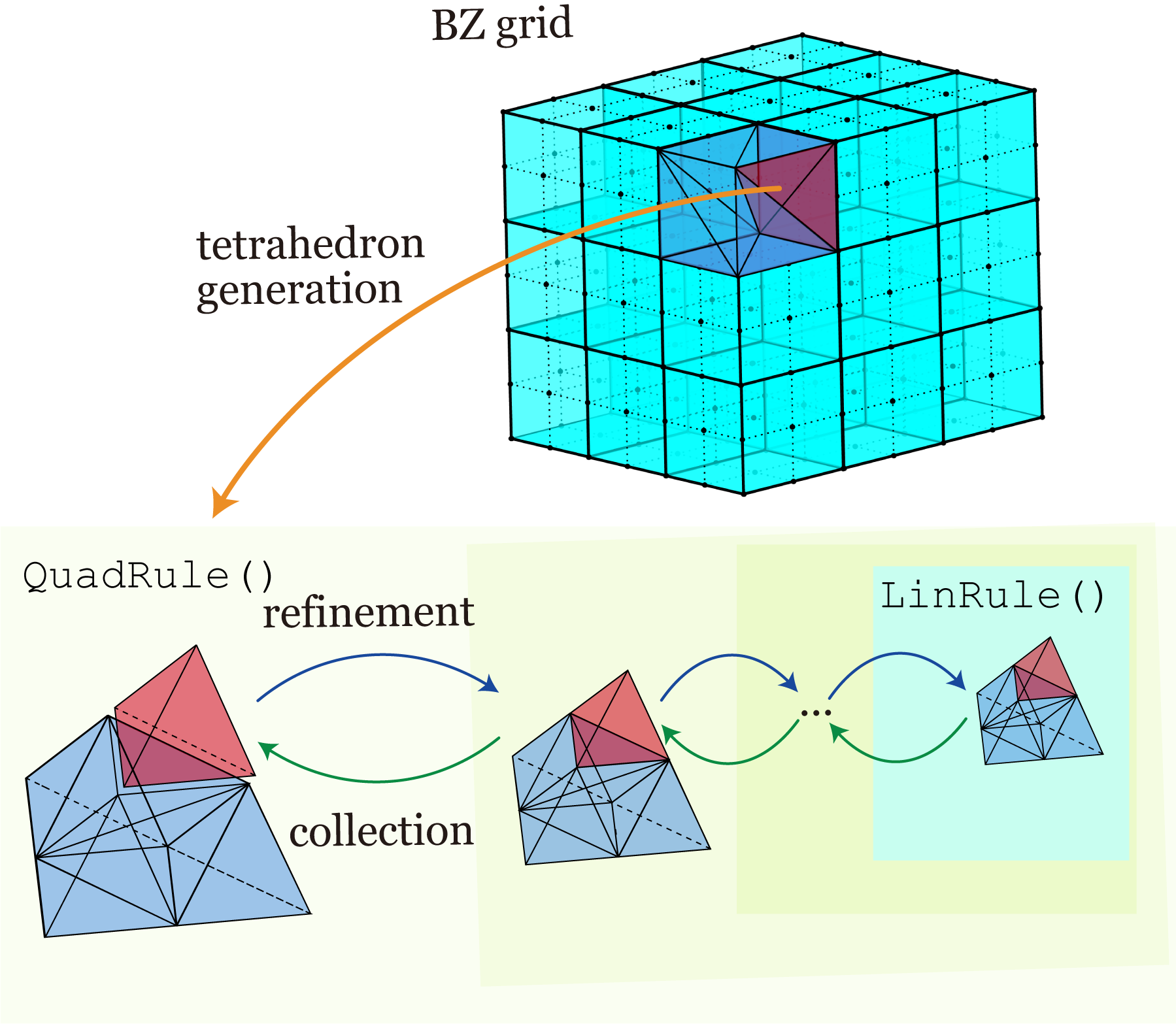}
    \caption{Overview of the recursive hybrid tetrahedron method.}
    \label{fig:rectet}
\end{figure}

Early in these developments, MacDonald \textit{et al.} proposed the \textit{hybrid tetrahedron method}, which refines a given $\vec k$-grid using quadratic interpolation and subsequently applies the linear tetrahedron method on the refined grid\cite{macdonald1979}. This approach reduces the error from the linear tetrahedron method without additional computation of the integrand, which is advantageous for time-consuming first-principles electronic structure calculations.
In this work, 
we extend the hybrid tetrahedron method into a recursive procedure that allows easy iterative tetrahedron refinements. The integral is expressed as a weighted sum over the initial $\vec k$-grid, and the weights are collected recursively after tetrahedron refinements, as schematically shown in Fig. \ref{fig:rectet}.
This method is shown to provide practical solutions for many BZ integrals encountered in realistic computation, including response function calculations in fairly sophisticated situations. The accuracy of our method should approach that of the quadratic tetrahedron method with sufficient iterative refinements. The paper is organized as follows: in Section \ref{HTM} an overview of the hybrid tetrahedron method is presented, where some of the key concepts and the motivation of our recursive hybrid tetrahedron method are discussed. This is followed in Section \ref{RHTM} by a description of the formalism and implementation of the recursive hybrid tetrahedron method. Finally, as demonstrations of the recursive tetrahedron method, Section IV presents three applications. The first application is a numerical calculation of the Lindhard function. The second is the calculation of the dynamical transverse spin susceptibility of the Hubbard model on a honeycomb lattice. The third application involves the calculation of the (bare) Kohn-Sham susceptibility for face-centered cubic (fcc) cobalt using density-functional theory (DFT)\cite{Kohn1965selconsistent}.

\section{hybrid tetrahedron method\label{HTM}}

In this section, we review the hybrid tetrahedron method\cite{macdonald1979} to prepare for the subsequent discussion of our recursive hybrid tetrahedron method. 

In all tetrahedron methods, the integration region is divided into a set of contiguous tetrahedra ($d$-simplices for dimension $d = 3$). This is usually achieved by setting up a regular $\vec k$-grid filling the parallelepiped BZ and then dividing each sub-cell into 6 tetrahedra (see Fig. \ref{fig:rectet}). 
Values at 4 non-coplanar points determine a unique linear form in $\mathbb{R}^3$. And given values on the $\vec k$-grid, the linear interpolants devised within each tetrahedron collectively form a continuous and piecewise linear function across the BZ.
The linear tetrahedron method exploits this linear interpolant of band eigenvalues to single out tetrahedra cut by the approximate Fermi surface, the occupied region of which are then split into sub-tetrahedra (see Appendix \ref{htsplit}). 
This way, the approximate occupied region is tiled with tetrahedra, over which the linear interpolant of the integrand may be analytically integrated.

The linear tetrahedron method converges rather slowly, at the quadratic order of characteristic $\vec k$-spacing $\Delta$. Two sources of error exist: overestimation (underestimation) of integral when integrating linear interpolant of a convex (concave) function, and the volume loss from tetrahedron tiling of the occupied region. Both errors can be reduced by the hybrid 
tetrahedron method\cite{macdonald1979}, which computes band eigenvalues and integrand on a particular refined $\vec k$-grid through quadratic interpolation and then applies the linear tetrahedron method. Upon repeated tetrahedron refinements, the error originating from the linear tetrahedron method can be made arbitrarily small. Then the final integral will converge to the result of the quadratic tetrahedron method, and the leading error now comes from the initial quadratic interpolation.

\begin{figure}[h]
    \centering
    \includegraphics[width=8cm]{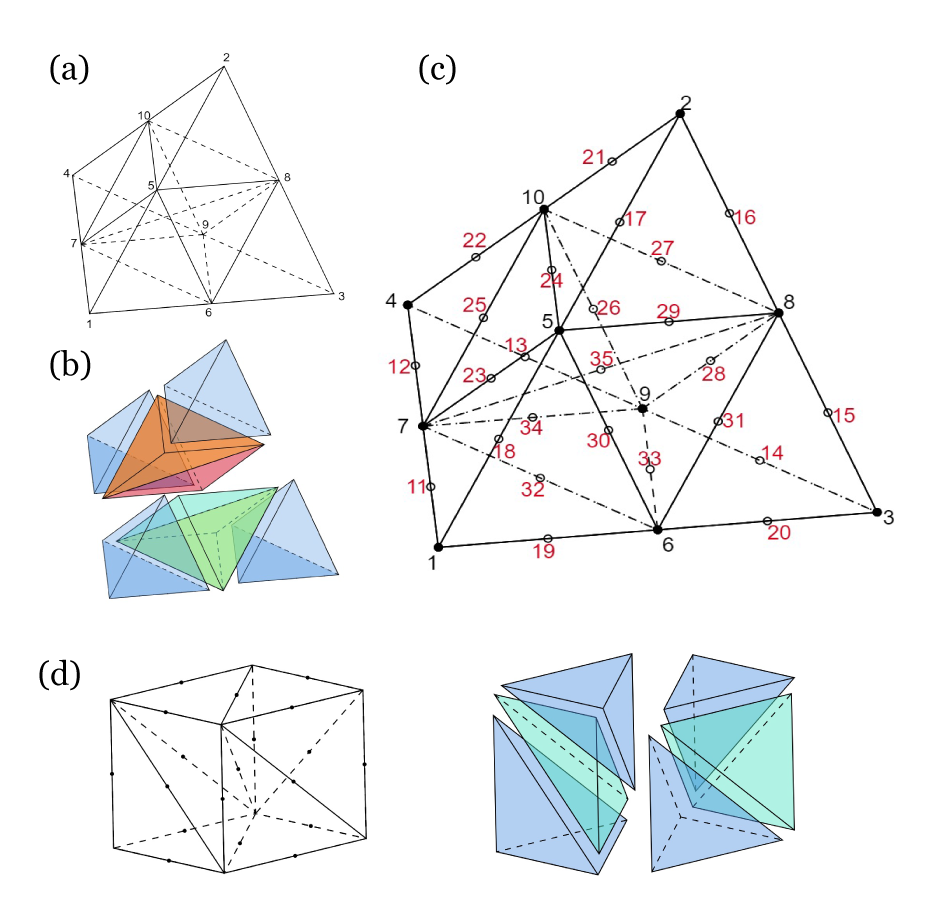}
    \caption{(a) Quadratic tetrahedron: a tetrahedron with function values at its vertices (labeled as $\vec k_{1-4}$) and edge midpoints (labeled as $\vec k_{5-10}$). (b) A quadratic tetrahedron can be divided into 8 isovolumetric sub-tetrahedra. (c) With their own edge midpoints (labeled as $\vec k_{11-35}$, red) added, each sub-tetrahedron can be made into a quadratic tetrahedron. Table. \ref{tab:coeff} follows the same label convention. (d) Division of a $2\times 2\times 2$ cube into 6 quadratic tetrahedra}
    \label{fig:QT}
  \end{figure}

Quadratic interpolation is compatible with tetrahedron methods because the function values at 4 vertices and 6 edge midpoints of a tetrahedron determine a unique quadratic form\footnote{In general, a unique quadratic form can be determined given values at 10 points in $\mathbb{R}^3$, if no more than 3 points are colinear and no more than 6 points are coplanar.}. Hereafter, we will refer to a tetrahedron with values at its 4 vertices (denoted by $\vec k_{1-4}$) and 6 edge midpoints (denoted by $\vec k_{5-10}$) as a ``quadratic tetrahedron" (see Fig. \ref{fig:QT}(a))
, while a ``linear tetrahedron" refers to a tetrahedron with function values at its 4 vertices only. 
As depicted in Fig. \ref{fig:QT}(b), a quadratic tetrahedron can be divided into 8 smaller linear tetrahedra of equal volumes. These linear tetrahedra can also be transformed into quadratic tetrahedra by computing additional function values at their own edge midpoints (denoted by $\vec k_{11-35}$ in Fig. \ref{fig:QT}(c)) as weighted sums of values at $\vec k_{1-10}$ (see Table. \ref{tab:coeff}) through quadratic interpolation. Like ordinary tetrahedron methods, we can divide the BZ into contiguous quadratic tetrahedra as well: set up a BZ-filling regular $\vec k$-grid, now with an even number of divisions in each dimension, and divide the BZ into contiguous $2\times 2\times 2$ sub-cells and then each sub-cell into 6 quadratic tetrahedra as shown in Fig. \ref{fig:QT}(d). Each quadratic tetrahedron can be further divided into 8 subordinate quadratic tetrahedra, whose vertices and edge midpoints together constitute a refined regular $\vec k$-grid with halved $\vec k$-spacing in each dimension.  
This division-into-eight process can be repeated so that the linear tetrahedron method on the final grid yields satisfactory accuracy.

\begin{table*}
    \caption{\label{tab:coeff}
    Quadratically interpolated values at $\vec k_{11-35}$ in Fig. \ref{fig:QT}(c) in terms of weighted sums of values at $\vec k_{1-10}$. Here, indices in parentheses represent values at the corresponding $\vec k$-points. For conciseness, the linear coefficients have been multiplied by 8. This table is adapted from Ref. \cite{macdonald1979} for readers' convenience.}
    \begin{ruledtabular}
    \begin{tabular}{cccc}
        \text{ Index } & \text{ Value } & \text{ Index } & \text{ Value }\\
        \hline
        11 & 6(7)+3(1)-(4) & 24 & 4(5)+4(10)+2(7)-(1)-(4)\\
        12 & 6(7)+3(4)-(1) & 25 & 4(7)+4(10)+2(5)-(1)-(2)\\
        13 & 6(9)+3(4)-(3) & 26 & 4(10)+4(9)+2(8)-(3)-(2)\\
        14 & 6(9)+3(3)-(4) & 27 & 4(10)+4(8)+2(9)-(4)-(3)\\
        15 & 6(8)+3(3)-(2) & 28 & 4(9)+4(8)+2(10)-(2)-(4)\\
        16 & 6(8)+3(2)-(3) & 29 & 4(5)+4(8)+2(6)-(1)-(3)\\
        17 & 6(5)+3(2)-(1) & 30 & 4(5)+4(6)+2(8)-(2)-(3)\\
        18 & 6(5)+3(1)-(2) & 31 & 4(6)+4(8)+2(5)-(1)-(2)\\
        19 & 6(6)+3(1)-(3) & 32 & 4(6)+4(7)+2(9)-(3)-(4)\\
        20 & 6(6)+3(3)-(1) & 33 & 4(6)+4(9)+2(7)-(1)-(4)\\
        21 & 6(10)+3(2)-(4) & 34 & 4(7)+4(9)+2(6)-(1)-(3)\\
        22 & 6(10)+3(4)-(2) & 35 & 2(5)+2(6)+2(7)+2(8)+2(9)\\
        23 & 4(5)+4(7)+2(10)-(2)-(4) &  & +2(10)-(1)-(2)-(3)-(4)\\
    \end{tabular}
    \end{ruledtabular}
    \end{table*}

Several remarks are in order, which highlight the motivation behind our recursive tetrahedron method. First, the quadratic interpolants in all quadratic tetrahedra together form a continuous and piece-wise quadratic function across the BZ. This is because the quadratic interpolant restricted to the shared face of a pair of adjacent quadratic tetrahedra (a 2-simplex) is uniquely determined by values at its vertices and edge midpoints (6 points in total). Second, the flexible tetrahedron refinement can be used to calculate response and spectral functions by quadratically interpolating both the numerator $A_{\vec k}$ and the denominator $D_{\vec k}$ on a refined $\vec k$-grid and computing the integral using Lambin and Vigneron's method\cite{lambin1984}. The hybrid tetrahedron method has a drawback. Both the integrand and the band eigenvalues undergo quadratic interpolation, and if we are to calculate multiple BZ integrals of the same type over the same band structure, we have to either store the interpolated band eigenvalues whose amount grows exponentially with tetrahedron refinements or recalculate them from scratch for each integral. In fact, this approach can be retooled into a recursive procedure, which expresses the integral as a weighted sum on the initial $\vec k$-grid and returns the weights as final results, freeing us from storing an enormous amount of interpolated values on a dense refined $\vec k$-grid. This is precisely the rationale for our recursive hybrid tetrahedron method, to be introduced next.

\section{Recursive hybrid tetrahedron method\label{RHTM}}

As pointed out in Bl\"ochl \textit{et al.}\cite{blochl1994}, an integral in the linear tetrahedron method can be transformed into a weighted sum over a $\vec k $-grid, where the weights are solely determined by band structure and can be calculated once and stored for reuse. In the hybrid tetrahedron method,
interpolated values of a function on a refined $\vec k$-grid are linear combinations of the evaluated values at the initial $\vec k$-grid. The linear coefficients depend solely on the geometric relationship between two grids.  
Thus, in principle, 
an integral in the hybrid tetrahedron method can also be transformed into a weighted
sum over the initial $\vec k$-grid and the integral weights can be calculated without explicit knowledge of the integrand.
In the following, we develop this idea into a universal formal framework, extend its applicability to more sophisticated BZ integrals with a variety of singularities, and present its implementation as a recursive procedure.

\subsection{Recursive formalism}
A generic BZ integral can be written as
\begin{equation}
I= \int\limits_{\text{BZ}}  W(\mathbf{k}) F(\mathbf{k})d\mathbf{k},
\end{equation}
where $F(\mathbf{k})$ is a regular function of $\mathbf{k}$, while the weight function $W(\mathbf{k})$ incorporates all the singularities of the integrand. Typical examples of $W(\mathbf{k})$ include
\begin{subequations}\label{Wex1}
    \begin{align}
    W(\mathbf{k}) & =  \Theta ( \varepsilon _{F} -\varepsilon (\mathbf{k}))\label{theta}\\
    W(\mathbf{k}) & =  \delta ( \varepsilon _{F} -\varepsilon \mathbf{( k}))\label{delta}\\
    W(\mathbf{k}) & =  \Theta ( \varepsilon _{F} -\varepsilon (\mathbf{k}))\frac{1}{D(\mathbf{k})}\label{thetafrac}\\
    W(\mathbf{k}) & =  \Theta ( \varepsilon _{F} -\varepsilon (\mathbf{k})) \delta(D(\mathbf{k}))\label{thetadelta} .
    \end{align}
\end{subequations}
The weight functions Eq. (\ref{theta}) and (\ref{delta}) are involved in the calculation of various physical quantities like charge density and density of states, while Eq. (\ref{thetafrac}) and (\ref{thetadelta}) are typically encountered in real and imaginary part of retarded response functions, respectively. Here we see three common types of singular factors: the step function, the Dirac-delta function, and a fraction with poles. Multiple singularities, often of different types, can appear simultaneously as in Eq. (\ref{thetafrac})\&(\ref{thetadelta}) and cause severe practical difficulties in numerical BZ integration.

In the hybrid tetrahedron method, one starts with an initial regular grid $g^{(0)}=\{\mathbf{k}^{(0)}_i\}$ filling the parallelepiped BZ, with an odd number of points separated by an even number of spaces in each dimension (periodic boundary condition is not assumed). 
Values of some smooth function $\phi(\mathbf{k})$ (e.g., $F(\vec k)$, $D(\vec k)$ and $\varepsilon(\vec k)$) are available on $g^{(0)}$, denoted $\phi^{(0)} =\{\phi(\mathbf{k}^{(0)}_i)\}$. The BZ can be divided into quadratic tetrahedra according to $g^{(0)}$. 
Applying the aforementioned divide-into-eight procedure to each quadratic tetrahedron leads to a refined grid $g^{(1)}=\{\mathbf{k}^{(1)}_i\}$ with half the spacings of $g^{(0)}$ in each dimension. The function values on $g^{(1)}$, denoted $\tilde \phi^{(1)}=\{\tilde{\phi}(\mathbf{k}^{(1)}_i)\}$, can be computed through quadratic interpolation according to Table. \ref{tab:coeff}. Here, the tilde distinguishes the true function $\phi(\mathbf{k})$ from its interpolant $\tilde{\phi}(\mathbf{k})$. 
This division process can be repeated, and the interpolated values on two successive grids are related by a linear map
\begin{equation}
    \tilde{\phi}(\mathbf{k}^{(p)}_i)=\sum_{j}Q_{ij}^{(p,p-1)}\tilde{\phi}(\mathbf{k}^{(p-1)}_j),
\end{equation}
which can be written succinctly as $\tilde \phi^{(p)}= Q^{(p,p-1)}\tilde \phi^{(p-1)}$.
We can then express the interpolated values on the final grid as a linear combination of the initial grid values
\begin{equation}
    \tilde{\phi}^{(N)}=Q^{(N,N-1)}\dots Q^{(2,1)}Q^{(1,0)}\phi^{(0)}
   \equiv Q^{(N,0)}\phi^{(0)},
\end{equation}
in which the composite linear map $Q^{(N,0)}$ is determined solely by the geometric relationship between $g^{(0)}$ and $g^{(N)}$.

As noted by Bl\"ochl \textit{et al.}\cite{blochl1994}, a numerical integral in any specialized linear tetrahedron method can be expressed as a weighted sum of $F(\vec k)$ over $g^{(N)}$:
\begin{equation}\label{LTonG}
   I \approx \sum_{i} w^{(N)}_i[W(\mathbf{k})]F(\mathbf{k}^{(N)}_i) .
\end{equation}
Note the integral weights $\{w^{(N)}_i\}$ are written as functionals
of $W(\mathbf{k})$, because the knowledge of $W(\mathbf{k})$ on $g^{(N)}$, usually in an approximated form, is needed to analytically integrate the linear interpolant of $F(\vec k)$ multiplied by $W(\vec k)$ within a linear tetrahedron. In the hybrid tetrahedron method, $\{F(\mathbf{k}^{(N)}_i)\}$ can be approximated by the interpolated values $\{\tilde{F}(\mathbf{k}^{(N)}_i)\}$, while $W(\vec k)$ contains singularities by definition and is not amenable to a direct interpolation. But in most practical cases, $W(\vec k)$ depends on $\mathbf{k}$ only through some smooth functions that can be properly interpolated. For example, the discontinuous step function $\Theta ( \varepsilon _{F} -\varepsilon (\mathbf{k}))$ depends on the band eigenvalue $\varepsilon (\mathbf{k})$ and the possibly divergent $1/D(\mathbf{k})$ depends on the smooth denominator $D(\mathbf{k})$. In these cases, $W(\vec k)$ on $g^{(N)}$ can be made available through quadratic interpolation of these interpolable functions. Finally we can approximate Eq. (\ref{LTonG}) as: 
\begin{equation}\label{main}
    \begin{aligned}
    I
    & \approx \sum_i w^{(N)}_i[W(\tilde{\varepsilon}(\mathbf{k}),\tilde{D}(\mathbf{k}),...)]\tilde{F}(\mathbf{k}^{(N)}_i)\\
    & \approx \sum_m\Big[\sum_i w^{(N)}_i[W(\tilde{\varepsilon}(\mathbf{k}),\tilde{D}(\mathbf{k}),...)]Q_{im}^{(N,0)}\Big]F(\mathbf{k}^{(0)}_m).
    \end{aligned}
\end{equation}
The result in the last line is written, with the help of ${Q}^{(N,0)}$, as a weighted sum of $\{F(\mathbf{k}^{(0)}_m)\}$ over the initial grid $g^{(0)}$, and terms in square bracket are identified as the integral weights being pursued.

Eq. (\ref{main}) is the main result of this work. In its present form, it is not restricted to the hybrid tetrahedron method with quadratic interpolation; any interpolation scheme that relates the interpolated values to the evaluated values by a linear map ${Q}^{(N,0)}$ can be employed with Eq. (\ref{main}).  However, the hybrid tetrahedron method has obvious advantages: local quadratic interpolation frees us from an explicit construction of ${Q}^{(N,0)}$, and the contributions from different quadratic tetrahedra can be independently (i.e., parallelizable) computed.

\subsection{Implementation}
Our implementation of the recursive hybrid tetrahedron method described above consists mainly of two algorithms, displayed in Algos. \ref{alg:rhtm} and \ref{alg:quad} below. Algo. \ref{alg:rhtm} is the primary subroutine, which divides the BZ into quadratic tetrahedra $QT=\{q\tau_i\}$ according to an initial grid $g^{(0)}$, calls a recursive function \texttt{QuadRule()} for each quadratic tetrahedron $q\tau_i$ to compute its contribution $\vec w_{i}$ to the integral weights $\{w(\vec k^{(0)}_i)\}$, and collects all the contributions back onto $g^{(0)}$. Algo. \ref{alg:quad} is the implementation of the recursive function \texttt{QuadRule()}. It iteratively divides the given quadratic tetrahedron $q\tau_i$ successively into subordinate quadratic tetrahedra for a given number of times and computes quadratically interpolated values at their additional edge midpoints. At the end of the iterative divisions,  function \texttt{LinRule()} applies a specialized linear tetrahedron method to each of the 8 (linear) sub-tetrahedra of every finest quadratic tetrahedron to compute their weights. All weights are subsequently collected  recursively back to the original quadratic tetrahedron $q\tau_i$ on the initial grid $g^{(0)}$ to generate $\vec w_{i}$. 

\SetAlgorithmName{Algo.}{algorithmautorefname}

\begin{algorithm}
    \caption{Recursive hybrid tetrahedron method}\label{alg:rhtm}
    \KwIn{$\{\varepsilon(\vec k_i^{(0)})\}$, $\{D(\vec k_i^{(0)})\}$ etc. on the initial $\vec k$-grid $g^0$, integer $iter$.} 

    \KwOut{Integral weights $\{w_{\vec k_i^{(0)}}\}$.}
    
    Divide $g^{0}$ into quadratic tetrahedra $QT=\{q\tau_i\}$.

    \For(){$\vec k_{1-10}=q\tau_i \in QT$}{
       $\vec{w}_i=\texttt{QuadRule}(\vec{\varepsilon}_i,\vec{D}_i,\dots,iter)$
       
    where $\vec{\phi}_i=[\phi(\vec k_1),\dots,\phi(\vec k_{10})]$, $\phi$ can be $\varepsilon,D$ etc. }

    \For(integral weights collection){$q\tau_i \in QT$}{
        \For {$\alpha = 1,2,...,10$}{
        $w_{\vec k_{i_\alpha}^{(0)}}+=\vec{w}_i[\alpha]$}

    $\vec k_{i_\alpha}^{(0)}$ in $g^0$ corresponds to $\vec k_{\alpha}$ in $q\tau_i$.
    }
\end{algorithm}

In each iteration within  \texttt{QuadRule()}, the current quadratic tetrahedron is referred to as $q\tau$, with its vertices, edge midpoints and additional edge midpoints of its subordinate quadratic tetrahedra denoted by $\vec k_{1-4}$, $\vec k_{5-10}$ and $\vec k_{11-35}$ as in the previous section. It could be divided into 8 linear tetrahedra $LT=\{l\tau_j\}$ with $l\tau_j=\{\vec k_{j_{1-4}}\}$, or 8 subordinate quadratic tetrahedra $sQT=\{q\tau_j\}$ with $q\tau_j=\{\vec k_{j_{1-10}}\}$. Indices $j_{1-10}$ for each $j$ are listed in Table. \ref{tab:subqt}, arranged with special care so that when $q\tau_j$ is treated as the new $q\tau$, the new $\vec k_{1-10}$ are the old $\vec k_{j_{1-10}}$ in correct order.

\begin{algorithm}
    \caption{QuadRule}\label{alg:quad}
    \SetKwFunction{FMain}{QuadRule}
    \SetKwProg{Fn}{Function}{:}{}
    \Fn{\FMain{$\vec{\varepsilon},\vec{D},\dots,iter$}}{
        \KwOut{Integral weights $\vec{w}$.}

        \textbf{Currently in} $\vec k_{1-10}=q\tau$.

    \eIf{$iter==0$}{
        Divide $q\tau$ into 8 linear tetrahedra $LT=\{l\tau_j\}$.
        $l\tau_j=\{\vec k_{j_{1}},\vec k_{j_{2}},\vec k_{j_{3}},\vec k_{j_{4}}\}$
        
        \For{$l\tau_j \in LT$} {
        $\vec{w}_j=\texttt{LinRule}(\vec{\varepsilon}_j,\vec{D}_j,\dots)$
        
            where $\vec{\phi}_j=[\vec{\phi}[j_1],\vec{\phi}[j_2],\vec{\phi}[j_3],\vec{\phi}[j_4]]$, $\phi$ can be $\varepsilon,D$ etc. 
            
        \For(weight collection){$\alpha=1,2,3,4$}{
                $\vec{w}[j_{\alpha}]+=\vec{w}_{j}[\alpha]$
            }
        }        
    }{
    Divide $q\tau$ into 8 subordinate quadratic tetrahedra $sQT=\{q\tau_j\}$.
$q\tau_j=\{\vec k_{j_{1}},\vec k_{j_{2}},\dots,\vec k_{j_{10}}\}$
    
    Quadratic interpolation at $\vec k_{1-35}$.
    $\bar{\vec \varepsilon}=\vec Q \cdot \vec \varepsilon ,\quad \bar{\vec D}=\vec Q \cdot \vec D ,\dots, $

    where $\bar{\vec{\phi}}=[\bar{\phi}[1],\bar{\phi}[2],\dots,\bar{\phi}[35]]$, $\phi$ can be $\varepsilon,D$ etc. $\vec Q$ is a $35\times 10$ matrix.

    \For(){$q\tau_j \in sQT$}
        {
            $\bar{\vec w}_j=\text{QuadRule}(\bar{\vec \varepsilon}_j,\bar{\vec D}_j,\dots,iter-1)$
            
        where $\bar{\vec \phi}_j=[\bar{\vec \phi}[j_1],\bar{\vec \phi}[j_2],\dots,\bar{\vec \phi}[j_{10}]]$ ($q\tau_j=\{\vec k_{j_{1}},\vec k_{j_{2}},\dots,\vec k_{j_{10}}\}$ in $\{\vec k_{1},\dots,\vec k_{35}\}$).

            \For(weight collection){$\alpha=1,2,\dots,10$}{
                $\bar{\vec w}[j_{\alpha}]+=\bar{\vec w}_{j}[\alpha]$
            }
        } 
        $\vec{w}=\vec Q^T \cdot \bar{\vec w}$
    }}
    \textbf{End Function}
\end{algorithm}

\begin{table}[htbp]
    \caption{\label{tab:subqt}%
    Indices of points of subordinate quadratic tetrahedra. Geometry is depicted in Fig. \ref{fig:QT}(c). The indices has been carefully arranged so that points $\vec k_{j_{1-10}}$ of master quadratic tetrahedron corresponds to $\vec k_{1-10}$ of $j-$th subordinate quadratic tetrahedron in correct order.}
    \begin{ruledtabular}
    \begin{tabular}{ccc}
    \text{ Index($j$) } & \text{ vertices($j_{1-4}$) } & \text{ edge midpoints ($j_{5-10}$) }\\
    \hline
    1 & 1,5,6,7 & 18,19,11;30,32,23\\
    2 & 5,2,8,10 & 17,19,24;16,27,21\\
    3 & 6,8,3,9 & 31,20,33;15,14,28\\
    4 & 7,10,9,4 & 25,34,12;26,13,22\\
    5 & 8,5,6,7 & 29,31,35;30,32,23\\
    6 & 5,7,8,10 & 23,29,24;35,27,25\\
    7 & 9,8,7,6 & 28,34,33;35,32,31\\
    8 & 7,10,9,8 & 25,34,35;26,28,27\\
    \end{tabular}
    \end{ruledtabular}
\end{table}

An unexplained piece in Algo. \ref{alg:quad} is the function \texttt{LinRule()}, which computes the weights of the linear tetrahedra from the final refinement.  \texttt{LinRule()} is crafted to handle a variety of singularities in $W(\vec k)$, including those in Eq. (2) and their combinations, which is capable of covering a wide range of applications in electronic structure, correlation function, and physical property calculations.

As mentioned in the previous section, there are three common singular factors, i.e. the step function, the Dirac-delta function, and a fraction with poles. The conventional linear tetrahedron method can handle the weight function Eq. (\ref{theta}), i.e. the step function\cite{jepson1971,lehmann1972}. A closed-form integral weights as functions of $\varepsilon_F$ for a single tetrahedron are available, which when differentiated against the parameter $\varepsilon_F$ yields the integral weights for Eq. (\ref{delta}), i.e. the Dirac-delta function.  

A full-fledged linear tetrahedron method\cite{lambin1984} for 
\begin{equation}\label{frac}
    W(\vec k) = \frac{1}{D(\mathbf{k})}
\end{equation}
for real-valued  $D(\vec k)$ has appeared in Lambin and Vigneron\cite{lambin1984} under the name "analytical tetrahedron method". But the formula of integral weights assume a specific denominator $D(\vec k)= z-\varepsilon(\vec k)$ and is still universal after proper substitution though a little bit awkward to use. We present an adaption in a more universal and symmetric form in Appendix. \ref{LTM}, along with the closed-form integral weights for Eq. (\ref{theta}) from Bl\"ochl \textit{et al}\cite{blochl1994}. 

Now we come to combining these basic linear tetrahedron methods to handle $W(\vec k)$ with multiple singular factors. Two important cases of composite singular factors  Eq. (\ref{thetadelta}) and (\ref{thetafrac}), encountered in response function calculations, both involve a step function.  We will first construct the tetrahedron-tiled occupied region on which the linear tetrahedron method for Eq. (\ref{delta}) and (\ref{frac}) can be applied, as described in Section \ref{intro}.   Integral contribution from each sub-tetrahedron
can be properly expressed as a weighted sum of values of $F(\vec k)$ at the vertices of the parent tetrahedron  (see Appendix. \ref{htsplit}). 

In principle, a function involving the product of step functions, such as
\begin{equation} \label{thetatheta}
    W(\mathbf{k})  = \Theta ( \varepsilon _{F} -\varepsilon _{1}(\mathbf{k})) \Theta ( \varepsilon _{F} -\varepsilon _{2}(\mathbf{k})),
\end{equation}
can be treated  by calculating the weights analyticaly for $\Theta ( \varepsilon _{F} -\varepsilon _{2}(\mathbf{k}))$ in every occupied tetrahedron corresponding to $\Theta ( \varepsilon _{F} -\varepsilon _{1}(\mathbf{k}))$, or the other way around. We take a different approach, in which by virtue of the identity
\begin{equation}
    \Theta (x_1) \Theta (x_2)=\frac{1}{2}[\Theta(x_1)+\Theta(x_2)-\Theta(-x_1x_2)]
\end{equation}
we impose the linear tetrahedron rule directly on the three step functions. It is acceptable to do so only when the tetrahedra on the final grid are sufficiently tiny, so that either $x_1x_2$ can be well approximated with 
a linear form
or the contributions from those tetrahedra in which both $x_1$ and $x_2$ change signs (and the leading order term of $x_1x_2$ becomes quadratic) can be neglected. 
Similarly, we can implement the linear tetrahedron method for
\begin{subequations}
    \begin{align}
    W(\mathbf{k}) & =  \Theta ( \varepsilon _{F} -\varepsilon _{1}(\mathbf{k})) \Theta ( \varepsilon _{F} -\varepsilon _{2}(\mathbf{k}))\delta(D(\mathbf{k}))\label{thetathetadelta}\\
    W(\mathbf{k}) & =  \Theta ( \varepsilon _{F} -\varepsilon _{1}(\mathbf{k})) \Theta ( \varepsilon _{F} -\varepsilon _{2}(\mathbf{k}))\frac{1}{D(\mathbf{k})}\label{thetathetafrac}
    \end{align}
\end{subequations}
by applying the method of Eq. (\ref{thetadelta}) and (\ref{thetafrac}) three times. 

Another common weight function is the joint density of states
\begin{equation}
    W(\mathbf{k})  = \delta ( \varepsilon _{F} -\varepsilon _{1}(\mathbf{k})) \delta ( \varepsilon _{F} -\varepsilon _{2}(\mathbf{k})),
\end{equation}
which occurs in, for example, Fermi surface nesting calculations\cite{kunes2004,johannes2008} or quasiparticle interference pattern\cite{markiewicz2004,mcelroy2006,simon2007a}. In our implementation, this is computed via  numerical differentiation by finite difference
\[\delta ( \varepsilon _{F} -\varepsilon _{1}) \delta ( \varepsilon _{F} -\varepsilon _{2})=\frac{\partial}{\partial\varepsilon}[\theta(\varepsilon-\varepsilon_1)\delta(\varepsilon_F-\varepsilon_2)]\Big|_{\varepsilon=\varepsilon_F}.\]

Finally, we would like to remark on incorporating the broadening method into our recursive hybrid tetrahedron method. This is more for simplified computation and rigorous physical correspondence than sampling convergence, which supposedly has been taken care of by iterative tetrahedron refinements. 
Typical examples include finite temperature calculations with occupation number $f_{\vec k}=1/(1+e^{\beta(\varepsilon_{\vec k}-\mu)})$ 
(finite $\beta$) and spectral function calculations with dissipative Green function $G_{\mathbf{k}}(\omega)=1/(\omega-\varepsilon_{\vec k}+i\delta)$ (finite $\delta$).  
Such calculations involve $F(\vec k)$ (the regular part of integrand) that contains structures difficult to resolve on a coarse $\vec k$-grid, e.g. the short tail of $f_{\vec k}$ near Fermi level at low temperatures and the narrow peak of $G_{\mathbf{k}}$ when $\omega\approx\varepsilon_{\vec k}$ in case of small relaxation time $\tau$ (note in case of $\beta\rightarrow \infty$ or infinitesimal $\tau$, $f_{\vec k}$ and $G_{\mathbf{k}}$ will be absorbed in $W(\vec k)$). Since the information of these small structures on a refined grid can still be made available by some slow varying auxiliary functions (e.g. $\varepsilon_{\vec k}$), we can approach them in the following manner. Take finite temperature calculations as an example:
\begin{eqnarray}
&& \int_{\text{BZ}}  W(\mathbf{k})[f( \varepsilon(\mathbf{k})) F'(\mathbf{k})]d\mathbf{k} \nonumber\\
    &\approx&\sum_i w^{(N)}_i [f( \varepsilon(\mathbf{k}_i^{(N)})) F'(\mathbf{k}_i^{(N)})] \nonumber\\
    &\approx& \sum_j \Big[\sum_i w^{(N)}_i f( \tilde{\varepsilon}(\mathbf{k}_i^{(N)}))Q_{ij}^{(N,0)}\Big] \tilde{F'}(\mathbf{k}_j^{(0)}).   
    \label{eq:finiteT} 
\end{eqnarray}

On the refined grid, we treat $f(\varepsilon(\vec k))F'(\vec k)$ as the regular part of the integrand and apply the linear tetrahedron method for $W(\vec k)$ to generate the integral weights $\{w^{(N)}_i\}$ on $g^{(N)}$. Then $f(\tilde \varepsilon(\vec k))$, with $\varepsilon(\vec k)$ replaced by its interpolant $\tilde{\varepsilon}(\vec k)$, is absorbed into the integral weights, which are then collected recursively back to the initial grid $g^{(0)}$. The terms in square brackets  in Eq. (\ref{eq:finiteT}) are identified as the integral weights on $g^{(0)}$. This approach is expected to work as long as the refined grid can well resolve the function structure of $F(\vec k)$ so that it can be approximated locally by a linear form.

\section{Applications\label{APP}}
In condensed matter physics, many experiments aim to measure the response of a system to external perturbations, which is often treated within the linear response theory\cite{kubo1957statistical}. 
Numerically, the BZ integrals of the linear response functions without exception involve multiple singularities (e.g., Eqs. \eqref{thetafrac} and \eqref{thetadelta}), requiring attention to the integration method and $\vec{k}$-grid settings. 
To compute the linear response functions by a direct summation over a $\vec{k}$-grid, a dense $\vec{k}$-grid combined with a substantial artificial broadening parameter ($\eta$) is often used. However, this direct summation approach can lead to artificial peaks due to numerical instability\cite{Muniz2002}, and an inappropriate broadening parameter may significantly alter the susceptibility structure\cite{Karlsson2000}. 
In previous studys of the tetrahedron methods, the linear tetrahedron method\cite{jepson1971,lehmann1972} and Bl\"ochl's improved linear tetrahedron method\cite{blochl1994} primarily address singularities arising from step functions, including those in charge density and density of states calculations. 
Lambin and Vigneron's linear analytical tetrahedron method\cite{lambin1984} is capable of analytically integrate the linear response functions in each tetrahedron but it converges with increasing $\vec{k}$-grid density at the same rate as other tetrahedron method using linear interpolation. 

Our recursive hybrid tetrahedron method is an extension of MacDonald's hybrid tetrahedron method\cite{macdonald1979} which combines quadratic interpolation and linear tetrahedron method. By incorporating iterative tetrahedron refinement and a novel recursive procedure, our recursive hybrid tetrahedron method following sufficient tetrahedron refinements is expected to achieve an accuracy comparable to that of the quadratic tetrahedron method. Additionally, we adapt Lambin and Vigneron's linear analytical tetrahedron approach\cite{lambin1984} to the integral on the refined $\vec{k}$-grid. This allows us to address multiple singularities while employing more accurate quadratic interpolation with much fewer sampling points. 

To demonstrate the advantages of our method mentioned above, we apply it to calculate linear respnse functions. 
To begin with, we consider the Lindhard function\cite{lindhard1954properties} which represents the density-density response of non-interacting electron gas, and compare the numerical results with analytical results. 
Subsequently, we calculate the dynamical transverse spin susceptibility of Hubbard model on a honeycomb lattice to extract magnon energies.  
Finally, we implement the recursive hybrid tetrahedron method in density functional theory\cite{kresse1996} to compute the Kohn-Sham susceptibility of fcc Co, enabling us to clearly identify regions of the Stoner continuum. 

\subsection{The Lindhard function}
The Lindhard function is a classic and representative example of linear response theory which has an analytical expressions\cite{giuliani2005}. 
Moreover, the response of an interacting system can be approximated by that of a non-interacting system under an effective self-consistent field, making the study of the density-density response function of a non-interacting electron gas an ideal starting point for analyzing interacting systems. 
The Lindhard function\cite{giuliani2005} in 3 dimensions writes 
\begin{equation}\label{lind}
    \chi_0(\vec q,\omega)=\frac{1}{(2\pi)^3}\int d^3 \vec{k} \frac{\Theta(\varepsilon_F-\varepsilon_{\vec k})-\Theta(\varepsilon_F-\varepsilon_{\vec k+\vec{q}})}{\varepsilon_{\vec k}-\varepsilon_{\vec k+\vec q}+\omega +i\eta}.
\end{equation}
Numerical calculation of the Lindhard function using broadening method requires a finite $\eta$ and the finite temperature Fermi distribution function with a smearing parameter $\sigma$ (as in $f_{\vec{k}}=\left[\exp\frac{{\varepsilon_{\vec{k}}-\varepsilon_{F}}}{\sigma}+1\right]^{-1}$) to avoid the singularities.
However, our recursive hybrid tetrahedron method for $W(\vec{k})=\frac{\Theta ( \varepsilon _{F} -\varepsilon_{\vec{k}})}{D(\mathbf{k})}$ and $W(\vec{k})=\Theta ( \varepsilon _{F} -\varepsilon_{\vec{k}}) \delta(D(\vec{k}))$ makes it possible to calculate the Lindhard functions numerically in the $\eta\rightarrow0^{+}$ and $T=0\,\mathrm{K}$ limit. 

\begin{figure}[htbp]
   \centering
   \includegraphics[width=8cm]{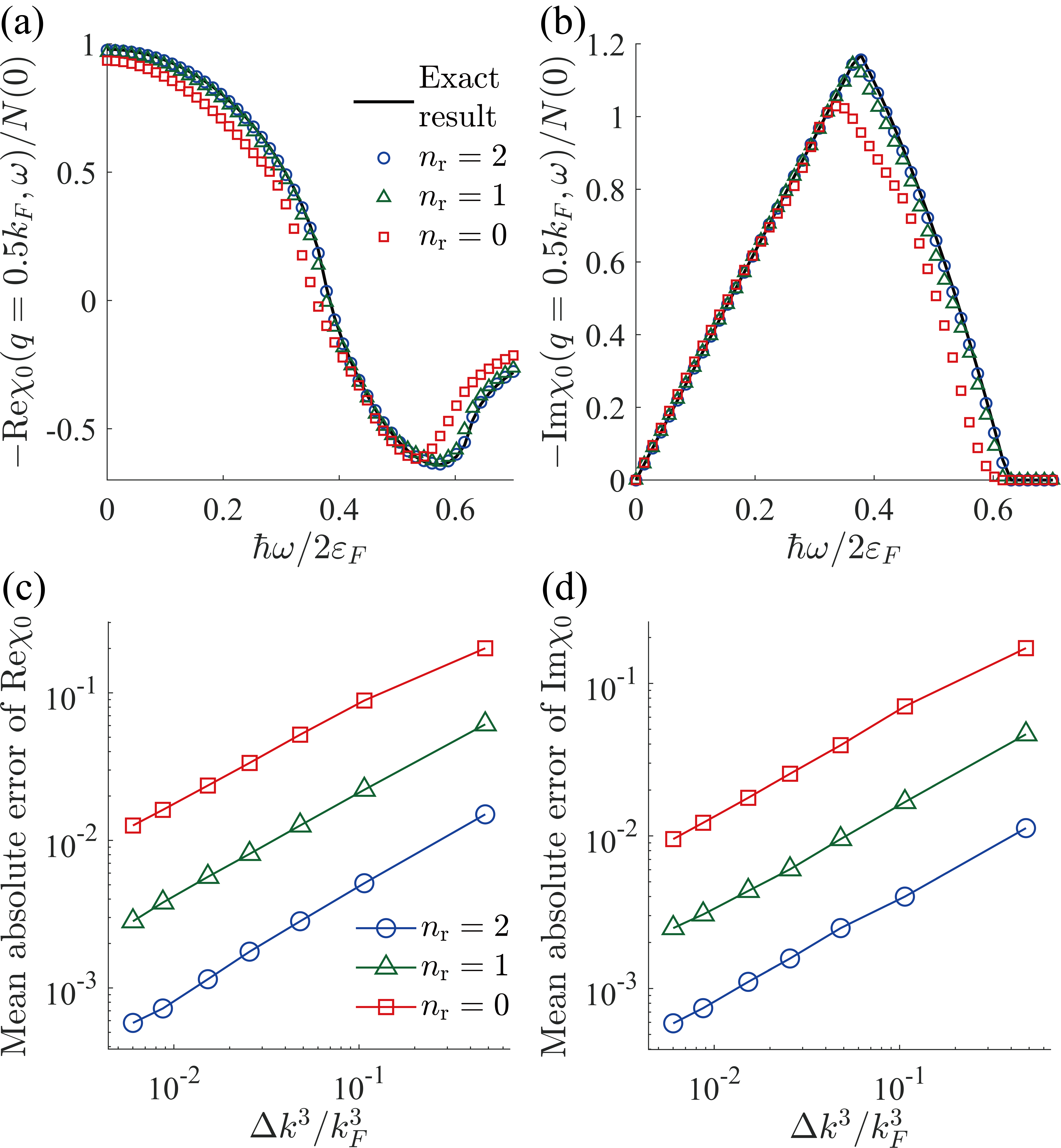}
   \caption{(a),(b)Real and imaginary part of $\chi_0(\vec q=0.5\vec{k}_F,\omega)$ calculated by the recursive hybrid tetrahedron method and exact analytical expression. $N(0)$ is the density of states at the Fermi surface. $k_F$ and $\varepsilon_F$ are Fermi vector and Fermi energy respectively.\
            (c),(d)Mean absolute error of $\mathrm{Re}\chi_0(\vec q=0.5\vec{k}_F,\omega)$ and $\mathrm{Im}\chi_0(\vec q=0.5\vec{k}_F,\omega)$ calculated by the recursive hybrid tetrahedron method with respect to the analytical result for different $\vec{k}$-spacings and $n_{\mathrm{r}}$. $\Delta\vec{k}^3$ is the volume that each $\vec{k}$-point occupies in reciprocal space.\
            Note that the recursive hybrid tetrahedron method with $n_{\mathrm{r}}=0$ is equivalent to the linear tetrahedron method, and $n_{\mathrm{r}}=1$ is equivalent to the original implementation of the hybrid tetrahedron method\cite{macdonald1979}. }
   \label{fig:Lindhard}
\end{figure}

In Fig. \ref{fig:Lindhard}, $n_{\mathrm{r}}$ represents the number of tetrahedron refinements in the recursive hybrid tetrahedron method. 
In Fig. \ref{fig:Lindhard}(a) and \ref{fig:Lindhard}(b), we used a relatively sparse $\vec{k}$-grid, with each $\vec{k}$-points occupying 0.11$k^{3}_{F}$ in the BZ. The linear tetrahedron method result ($n_{\mathrm{r}}=0$) shows a noticeable deviation from the exact result. 
The recursive hybrid tetrahedron method with one tetrahedron refinement ($n_{\mathrm{r}}=1$) is equivalent to the hybrid tetrahedron method\cite{macdonald1979}. A single refinement achieves improved accuracy over the linear tetrahedron method, though a second refinement is seen to result in further improvement in the accuracy. 
Fig. \ref{fig:Lindhard}(c) and (d) illustrate that the mean absolute error (MAE) for different $n_\mathrm{r}$ values is log-log linear with respect to integration volume element $\Delta k^3$. More important, the MAE is seen to decrease exponentially with the number of tetrahedron refinements, demonstrating clear advantage of the recursion in our method. 

\subsection{Dynamical transverse susceptibility of Hubbard model on the honeycomb lattice}    
The Hubbard model is very important in studying electron correlation and magnetism in condensed matter physics\cite{hubbard1963electron,kanamori1963electron,Gutzwiller1963,tasaki1998hubbard}. To investigate the spin dynamics for the Hubbard model, we compute the dynamical transverse spin susceptibility $\mathrm{Im}\chi_{+-}(\boldsymbol{q},\omega)$ under the random phase approximation (RPA)\cite{tang1998theory,Muniz2002}. 
The magnon energies can be extracted from the peaks in $\mathrm{Im}\chi_{+-}(\boldsymbol{q},\omega)$. 

The Hubbard model on a 2-dimensional honeycomb lattice is considered. For simplicity, each lattice site contains a single orbital, and our focus is on the half-filled antiferromagnetic phase. The Hamiltonian for the Hubbard model reads
\begin{equation}
    H=H^{0}+H^{U}=-t\sum_{\langle i,j\rangle\sigma}c_{i\sigma}^\dagger c_{j\sigma}+U\sum_{i} n_{i\uparrow}n_{i\downarrow},
\end{equation}
where $i$ and $j$ label lattice sites, $\sigma$ denotes spin, and the operators $c$ and $n$ represent the electron annihilation and number operators, respectively. The notation $\langle i,j\rangle$ refers to nearest neighbors on the honeycomb lattice. Here, the hopping term $H^{0}$ includes only nearest-neighbor hopping $t$, while the second term represents the intraorbital Hubbard interaction. 
Under the Hartree-Fock approximation, we Fourier-transform the Hubbard model Hamiltonian into momentum space and diagonalize it as $H=\sum_{n\boldsymbol{k}}E_{n\boldsymbol{k}}d_{n\boldsymbol{k}}^\dagger d_{n\boldsymbol{k}}$, where $d_{n\boldsymbol{k}}^\dagger=\sum_{l\sigma'} a_{n}^{l\sigma'}(\boldsymbol{k})c_{l\boldsymbol{k}\sigma'}^\dagger$ with $l=A,B$ denoting the sub-lattices. Previous studys on the Hubbard model for a honeycomb lattice\cite{Raczkowski2020,sorella1992semi} have identified a metal-insulator transition at $U_{c}/t\approx2.23$ using mean-field theory. In this study, we select $U/t=3$, where the mean-field theory gives an antiferromagnetic insulator. 
The bare transverse susceptibility is expressed as
\begin{equation}
    \begin{aligned}
        \left(\chi^0_{+-}\right)_{l_3l_4}^{l_1l_2}&(\boldsymbol{q},\omega)=-\frac{\Omega}{(2\pi)^2}\int d\boldsymbol{k}\sum_{nn^{\prime}}\left(f_{n\boldsymbol{k}}-f_{n^{\prime}\boldsymbol{k}+\boldsymbol{q}}\right)\\
        &\times\frac{a_{n^{\prime}}^{l_2\downarrow}(\boldsymbol{k}+\boldsymbol{q})a_{n^{\prime}}^{l_3\downarrow*}(\boldsymbol{k}+\boldsymbol{q})a_{n}^{l_4\uparrow}(\boldsymbol{k})a_{n}^{l_1\uparrow*}(\boldsymbol{k})}{\omega+E_{n\boldsymbol{k}}-E_{n^{\prime}\boldsymbol{k}+\boldsymbol{q}}+i\eta},
        \label{chi0+-}
    \end{aligned}
\end{equation}
where $\Omega$ is the area of the primitive cell and $\eta$ is set to be $t/20$. 
In the RPA, the transverse susceptibility becomes 
\begin{equation}
    \chi_{+-}=(1-\chi^{0}_{+-}U^s)^{-1}\chi^{0}_{+-},
\end{equation} 
where $U^s$ represents the interaction matrices. The only nonzero matrix element of $U_{s}$ is $(U^s)^{ll}_{ll}=U$.

\begin{figure}[htbp]
    \centering
    \includegraphics[width=8.3cm]{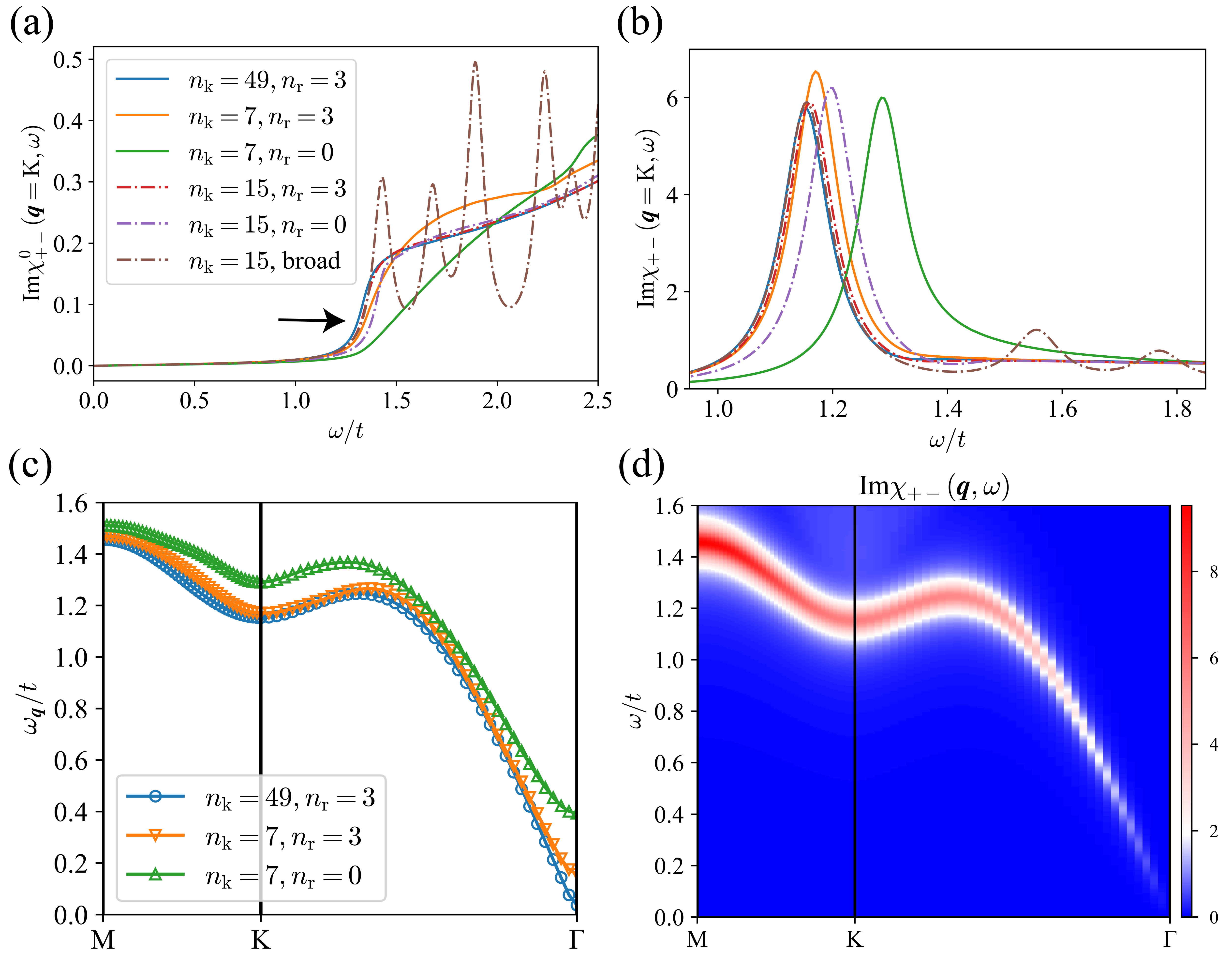}
    \caption{(a), (b)$\mathrm{Im}\chi^{0}_{+-}(\boldsymbol{q},\omega)$ and $\mathrm{Im}\chi_{+-}(\boldsymbol{q},\omega)$ as a function of $\omega$ for $\boldsymbol{q}=\mathrm{K}$. (c)The magnon energy dispersion for different $n_{\mathrm{k}}$ and $n_{\mathrm{r}}$. $n_{\mathrm{k}}$ means that we use a $n_{\mathrm{k}}\times n_{\mathrm{k}}$ $\vec{k}$-grid. (d)The spin fluctuation spectrum $\mathrm{Im}\chi_{+-}(\boldsymbol{q},\omega)$ ($n_{\mathrm{k}}=49$, $n_{\mathrm{r}}=3$).} 
    \label{fig:hubbard}
 \end{figure}

In Fig. \ref{fig:hubbard}, we compare the dynamical transverse susceptibility and magnon energies calculated using the recursive hybrid tetrahedron method, the linear tetrahedron method and the broadening method (direct summation). Here, $n_{\mathrm{k}}$ denotes the number of $\vec{k}$-points per dimension in the $\Gamma$-centered $\vec{k}$-grid, with "$n_{\mathrm{k}}=49,n_{\mathrm{r}}=3$" used as a reference following convergence testing. In Fig. \ref{fig:hubbard}(a), the nonzero $\mathrm{Im}\chi^{0}_{+-}(\boldsymbol{q},\omega)$ indicates the presence of the Stoner continuum, where the singularities in the denominators of \eqref{chi0+-} are particularly severe. The broadening method fails to converge at a $15\times15$ $\vec{k}$-grid, and numerous artifacts including wiggles are observed, which similarly occur in calculations for the real material fcc Co to be discussed in the following subsection. The recursive hybrid tetrahedron method ($n_{\mathrm{k}}=15,n_{\mathrm{r}}=3$) yields results almost identical to the reference, whereas the linear tetrahedron method ($n_{\mathrm{k}}=15,n_{\mathrm{r}}=0$) still shows some deviation, particularly in the region indicated by the arrow in Fig. \ref{fig:hubbard}(a). With a coarse $7\times7$ $\vec{k}$-grid, the recursive hybrid tetrahedron method with tetrahedron refinements ($n_{\mathrm{3}}$) already improves the result and removes the artificial peaks. Fig. \ref{fig:hubbard}(b) shows the dynamical transverse susceptibility at $q=K$ as a function of $\omega$, with a magnon peak. When using a $7\times7$ $\vec{k}$-grid, the linear tetrahedron method ($n_{\mathrm{r}}=0$) overestimates the magnon energy due to insufficient $\vec{k}$-grid density for accurate band structure representation via linear interpolation. 
In contrast, our recursive hybrid tetrahedron method, which incorporates quadratic interpolation ($n_{\mathrm{r}}=3$ in Fig. \ref{fig:hubbard}), achieves results comparable to the reference even with a coarse $7\times7$ $\vec{k}$-grid. A finer $15\times15$ $\vec{k}$-grid brings the results coloser to the reference. Nevertheless, the $15\times15$ $\vec{k}$-grid with $n_{\mathrm{r}}=0$ still slightly overestimates the magnon energy, and the broadening method produces artificial peaks alongside the magnon excitation, complicating the determination of magnon energies. 
We used Lorentzian fitting to extract the magnon energy dispersion from the dynamical transverse susceptibility\cite{liu2023}. 

As shown in Fig. \ref{fig:hubbard}(c), the linear tetrahedron mehtod introduces a pronounced "Goldstone gap" and consistently overestimates the magnon energy along the high-symmetry $\vec{k}$-path, indicating that this overestimation is not limited to specific $\vec{k}$-points. The recursive hybrid tetrahedron method with tetrahedron refinements reduces this overestimation error. Finally, Fig. \ref{fig:hubbard}(d) presents the overall spin fluctuation spectrum ($n_{\mathrm{k}}=49,n_{\mathrm{r}}=3$), which clearly demonstrates the linear dispersion characteristic of an antiferromagnetic system in the long-wavelength limit. 

\subsection{Kohn-Sham susceptibility of fcc Co}
In addition to numerical calculations based on model Hamiltonians, our recursive hybrid tetrahedron method can also be implemented within the Kohn-Sham density-functional theory (DFT)\cite{Kohn1965selconsistent}. 
Here, we focus on the first-order perturbation approach in time-dependent (TD) DFT\cite{Runge1984density}, which allows for the calculation of dynamical responses to external electromagnetic field. 
However, evaluating BZ integrals for response functions within TD DFT presents significant challenges due to the complexity of first-order perturbation theory\cite{liu2023}. 

For a comprehensive discussion of the formalism underlying first-order perturbation in TD DFT, we refer readers to \cite{liu2023}, and we adopt the same notation in this work. The primary step in our implementation is solving for the first-order wavefunctions. Given that we are working with a crystalline solid, it is convenient to decompose physical quantities into different channels. We label each channel by $\ell\equiv(\nu,\vec{l},\vec{h})$. 
Defining $H$ as the ground-state Kohn-Sham Hamiltonian, $\delta H$ as the first-order Hamiltonian, and $H_{\vec{k}}=e^{-i\vec{k}\cdot\vec{r}}He^{i\vec{k}\cdot\vec{r}}$ as the Bloch Hamiltonian, the first-order perturbation equations in the $\ell$ channel are expressed as\cite{liu2023}
\begin{equation}
    (\nu+i\eta+\varepsilon_{n\vec{k}}-H_{\vec{k}+\vec{l}})\ket{u_{n\vec{k}}^{(1)}(\ell)}=\delta H(\vec{r},\ell)\ket{u_{n\vec{k}}},
    \label{Sternheimer}
\end{equation}
where $\ket{u_{n\vec{k}}^{(1)}(\ell)}$ and $\ket{u_{n\vec{k}}}$ represent the cell-periodic components of the first-order and ground-state wavefunctions, respectively. 
From the formal solutions of Eq. \eqref{Sternheimer}
$
    \ket{u_{n\vec{k}}^{(1)}(\ell)}=\sum_{n^{\prime}}\frac{\ket{u_{n^{\prime}\vec{k}+\vec{l}}}\bra{u_{n^{\prime}\vec{k}+\vec{l}}}\delta H(\vec{r},\ell)\ket{u_{n\vec{k}}}}{\nu+\varepsilon_{n\vec{k}}-\varepsilon_{n^{\prime}\vec{k}+\vec{l}}+i\eta}
$,
we can calculate the first-order density change
$
    \Delta\vec{\rho}(\ell)=\sum_{n\vec{k}}f_{n\vec{k}}\left(\bra{u^{(1)}_{n\vec{k}}(-\ell)}\vec{\sigma}\ket{u_{n\vec{k}}}+\bra{u_{n\vec{k}}}\vec{\sigma}\ket{u^{(1)}_{n\vec{k}}(\ell)}\right)
$
with the explicit form
\begin{equation}
    \begin{aligned}
        \Delta\vec{\rho}(\ell)
        =&\frac{\Omega}{(2\pi)^3}\int d\vec{k}\sum_{n^{\prime},n}f_{n\vec{k}}\times\\
        &\left(\frac{\bra{u_{n\vec{k}}}\delta H^{\dagger}(\vec{r},-\ell)\ket{u_{n^{\prime}\vec{k}-\vec{l}}}\bra{u_{n^{\prime}\vec{k}-\vec{l}}}\vec{\sigma}\ket{u_{n\vec{k}}}}{-\nu+\varepsilon_{n\vec{k}}-\varepsilon_{n^{\prime}\vec{k}-\vec{l}}-i\eta}\right.\\
        &\left.+\frac{\bra{u_{n\vec{k}}}\vec{\sigma}\ket{u_{n^{\prime}\vec{k}+\vec{l}}}\bra{u_{n^{\prime}\vec{k}+\vec{l}}}\delta H(\vec{r},\ell)\ket{u_{n\vec{k}}}}{\nu+\varepsilon_{n\vec{k}}-\varepsilon_{n^{\prime}\vec{k}+\vec{l}}+i\eta}\right),\\
    \end{aligned}
\end{equation}
where $\Omega$ is the volume of the primitive cell. 

In practical calculations with broadening method, achieving convergence often requires a very dense $\vec k$-grid (e.g., $50\times 50\times 50$ $\vec k$-grid as in Ref. \cite{cao2018}) due to strongly localized structures in the integrand. 
Both terms in Eq. (\ref{Deltarho}) contain denominators that can approach zero. Since the principal integrals are explicit, we apply our recursive hybrid tetrahedron method for $W(\vec k)=1/D(\mathbf{k})$, treating $(\nu+\varepsilon_{n\vec{k}}-\varepsilon_{n^{\prime}\vec{k}+\vec{l}}+i\eta)^{-1}$ as the singular part. 
After generating integral weights for each term in Eq. (\ref{Deltarho}), we compute the first-order density change using
\begin{equation}
    \begin{aligned}
    &\Delta\vec{\rho}(\ell)=\sum_{n^{\prime},n,\vec{k}}f_{n\vec{k}}\\
    &\times\left(W_{nn^{\prime}\vec{k}}(\ell)\bra{u_{n\vec{k}}}\vec{\sigma}\ket{u_{n^{\prime}\vec{k}+\vec{l}}}\bra{u_{n^{\prime}\vec{k}+\vec{l}}}\delta H(\vec{r},\ell)\ket{u_{n\vec{k}}}\right.\\
    &\left. +W_{nn^{\prime}\vec{k}}(-\ell)\bra{u_{n\vec{k}}}\delta H^{\dagger}(\vec{r},-\ell)\ket{u_{n^{\prime}\vec{k}-\vec{l}}}\bra{u_{n^{\prime}\vec{k}-\vec{l}}}\vec{\sigma}\ket{u_{n\vec{k}}}\right).
    \end{aligned}
    \label{Deltarho}
\end{equation}
Note that $W_{nn^{\prime}\vec{k}}(\ell)$ and $W_{nn^{\prime}\vec{k}}(-\ell)$ are not complex conjugate of each other. 
The TD DFT code\cite{liu2023} is based on the Vienna \textit{ab initio} simulation package (VASP)\cite{kresse1996}. 
For our calculations, we use an experimental lattice parameter of 3.548\,\AA\cite{wyckoff1963cubic}, a $\Gamma$-centered $\vec{k}$-grid and a plane-wave energy cutoff of 500 eV. 
In the linear response calculations, the broadening parameter $\eta$ is set to 10 meV. 

\begin{figure*}[htbp]
    \centering
    \includegraphics[width=17cm]{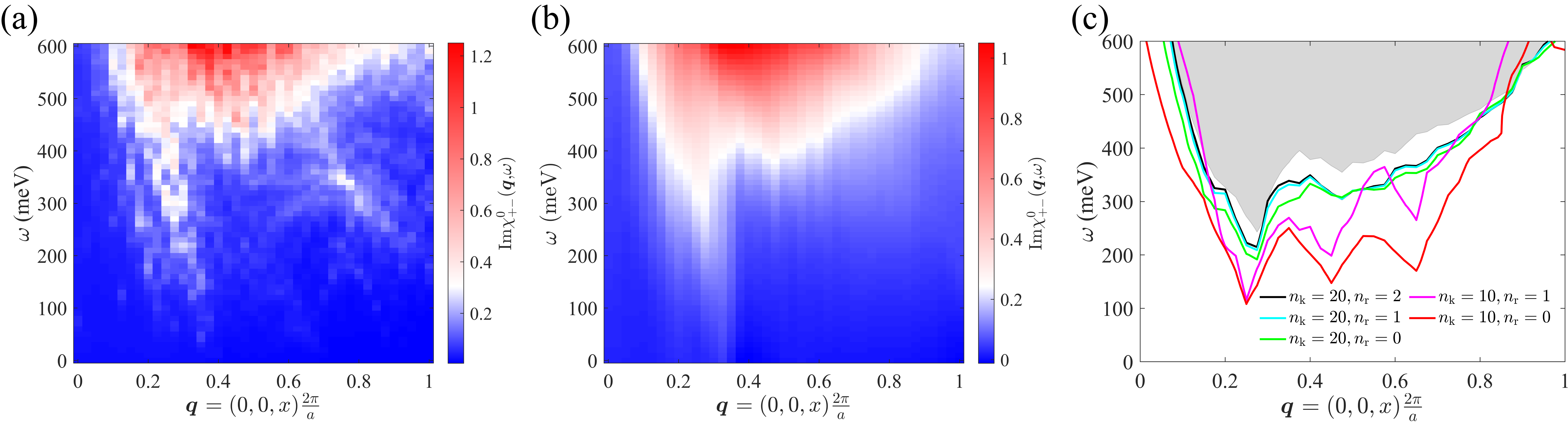}
    \caption{$\mathrm{Im}\chi^{0}_{+-}(\vec{q},\omega)$ of fcc Co along the [001] directions calculated using (a) the broadening (smearing) method and (b) the recursive tetrahedron method. (c) Comparison of boundary of the Stoner continuum calculated by the recursive tetrahedron method with different $\vec{k}$-mesh and iteration numbers.}
    \label{fig:Cochi0}
\end{figure*}

After obtaining the Kohn-Sham susceptibility spectrum, we can identify the Stoner continuum region, where spin waves can undergo Landau damping\cite{landau1946vibrations}. This information is valuable for both experimental and theoretical investigation on spin excitations. 
Fig. \ref{fig:Cochi0}(a) and \ref{fig:Cochi0}(b) show the Kohn-Sham susceptibility of fcc Co along the $\Gamma-\mathrm{X}$ $\vec{k}$-path (i.e., [001] direction) calculated with a $30\times30\times30$ $\vec{k}$-grid. 
Fig. \ref{fig:Cochi0}(a) presents the result obtained with the broadening method, while Fig. \ref{fig:Cochi0}(b) shows the result from the recursive tetrahedron method with a single tetrahedron refinements. 
It is evident that the recursive hybrid tetrahedron method produces sensible  results, whereas the broadening method produces unexpected jaggedness in view of the band structure. 

Next, we determine the Stoner continuum boundaries on $q_z$-$\omega$ plane for cobalt, using different $\vec{k}$-grids and $n_\mathrm{r}$'s as shown in Fig. \ref{fig:Cochi0}(c). 
For $\mathrm{Im}\chi^{0}_{+-}(\vec{q},\omega)$ over the plotted $\vec{q}$ and $\omega$ range, we normalize $\mathrm{Im}\chi^{0}_{+-}(\vec{q},\omega)$ based on its maximum and minimum values, defining 0.2 as the threshold for the Stoner continuum boundary. 
The calculation with a $30\times30\times30$ $\vec{k}$-grid and a single refinement serves as the reference, with the Stoner continuum shown as a gray area. With a $20\times20\times20$ $\vec{k}$-grid, we can see that with zero, one or two tetrahedron refinements the boundaries improve consistently towards the reference. 
The same trend can be seen with a $10\times10\times10$ $\vec{k}$-grid. 

\section{Summary and concluding remark}
We have extended the hybrid tetrahedron method to a recursive procedure, supplementing it with iterative tetrahedron refinements followed by an integral weights collection. In this way, a numerical BZ integral is transformed into a weighted sum over the initial $\vec k$-grid, and the quality of integral weights returned by our method approaches that of the direct quadratic tetrahedron method, featuring considerably superior convergence compared with the linear tetrahedron method. Iterative tetrahedron refinements also allow for flexible strategies to simultaneously handle multiple singularities in integrands and make our method applicable to a broad range of BZ integrals arising in practical calculations, including particularly the calculation of response and spectral functions. We have demonstrated the excellent performance of our method through two direct numerical integrations of the density response functions for model Hamiltonians and a TD DFT calculation of the Kohn-Sham susceptibility for fcc Co. 

We also provide a Julia implementation of the recursive hybrid tetrahedron method\cite{bzint}. Several matters need attention in the application of our recursive hybrid tetrahedron method. First, the open boundary condition, rather than the usual periodic boundary condition, is assumed in the working $\vec k$-grid. Thus, a $9\times 9\times 9$ $g^0$ will correspond to an $8\times 8\times 8$ $\vec k$-grid in case of the periodic boundary. We choose this convention because, unlike Fourier interpolation, quadratic interpolation does not presume periodicity and depends on the local information of the interpolant. Second, crystal symmetries are widely used to confine $\vec k$-sampling within an irreducible wedge of the BZ in modern electronic structure calculations. In order for the recursive hybrid tetrahedron method (or any modern tetrahedron method) to work, quantities over the whole BZ must be recovered by proper symmetry transformations. 

The recursive hybrid tetrahedron method can be applied in full self-consistent density functional perturbation theory (DFPT)\cite{liu2023}. In this method, the BZ integral is expressed as a weighted summation over the initial $\vec{k}$-grid, with integral weights calculated once for a given electronic band structure. Reusing these integral weights within the self-consistent DFPT loop is expected to increase the accuracy of the entire loop through quadratic interpolation and analytical tetrahedron integration, without incurring additional computational costs for recalculating integral weights. 
However, an unsettled issue is the proper treatment of multi-band systems, a problem that permeates but seems overlooked in the modern application 
of many tetrahedron methods based on analytical integration. The problem is that improper band labeling can cause uncertain discontinuities in band eigenvalue $\varepsilon_{n\vec{k}}$ and other band-dependent quantities as functions of $\vec k$. In most modern electronic structure software, band eigenvalues at each $\vec k$ are output in a sorted sequence with the band index $n$ only indicating their sizes, and thus $\varepsilon_{n\vec{k}}$ becomes discontinuous in case of band crossing. This issue for the linear tetrahedron method of Eq. (\ref{theta}) is not very severe since integral weights become non-trivial only near the Fermi level, and local smoothness only  within those tetrahedra crossed by the Fermi surface is required.
However, other linear tetrahedron methods, e.g., of Eq. (\ref{delta}) and Eq. (\ref{frac}), will be seriously affected. Higher-order interpolations, like quadratic interpolation used in the hybrid tetrahedron method, usually cause overfitting in the case of discontinuous functions, which may make the situation worse. Thus, for our recursive hybrid tetrahedron method to work at its full potential, an effective band sorting algorithm that restores the continuity in $\varepsilon_{n\vec{k}}$ is needed.

\section{ACKNOWLEDGMENTS}
We acknowledge the financial support from the National Natural Science Foundation of China (Grant Nos. 12274003, 11725415 and 11934001), the National Key R\&D Program of China (2021YFA1400100), and the Innovation Program for Quantum Science and Technology (Grant No. 2021ZD0302600). 

\begin{widetext}
\section{Appendix : linear tetrahedron method\label{LTM}}
In this Appendix, we summarize the linear tetrahedron methods for Eq.(\ref{theta}),(\ref{delta}) and (\ref{frac}). The closed-form linear tetrahedron rules for Eq.(\ref{theta}) is adopted from Ref.\cite{blochl1994} for reader's convenience, but are presented as a pedagogical example to show how to handle $\Theta ( \varepsilon _{F} -\varepsilon (\mathbf{k}))$ in the weight function $W(\vec k)$.
They when differentiated against the parameter $\varepsilon_F$ yield the rules for Eq.(\ref{theta}). The rules for Eq.(\ref{frac}) are derived by us independently and are cast in a symmetric form that facilitates implementation. A full-fledged linear tetrahedron method for Eq.(\ref{frac}) can be also found in Ref.\cite{lambin1984}, though designed for a specific $D(\vec k)= z-\varepsilon(\vec k)$.

\subsection{Split a tetrahedron according to \texorpdfstring{$\Theta ( \varepsilon _{F} -\varepsilon (\mathbf{k}))$}{Split a tetrahedron according to Θ(εF−ε(k))}\label{htsplit}}

As mentioned in section \ref{HTM}, the linear tetrahedron method handles factor $\Theta ( \varepsilon _{F} -\varepsilon (\mathbf{k}))$ by cutting each tetrahedron crossed by the approximated Fermi surface, a constant plane of the linear interpolant of $\varepsilon(\mathbf{k})$ at $\varepsilon _{F}$, and further splitting its occupied part into sub-tetrahedra. There are three cases depending on how the Fermi surface crosses the tetrahedron as shown in Fig.\ref{fig:split}, each with a specific treatment.  Hereafter, we assume the vertices of the tetrahedron are labeled as $\vec k_{1-4}$ according to the order of their band eigenvalues $\varepsilon_{1-4}$, i.e. $\varepsilon _{1} \leqslant \varepsilon _{2} \leqslant \varepsilon _{3} \leqslant \varepsilon _{4}$. The volume of the tetrahedron is denoted $V_T$. Note in the case $\varepsilon _{3} < \varepsilon _{F} \leqslant \varepsilon _{4}$, we could have split the occupied part but didn't, since the unoccupied part itself is already a sub-tetrahedron. 

\begin{figure}[H]
    \centering
    \includegraphics[width=8cm]{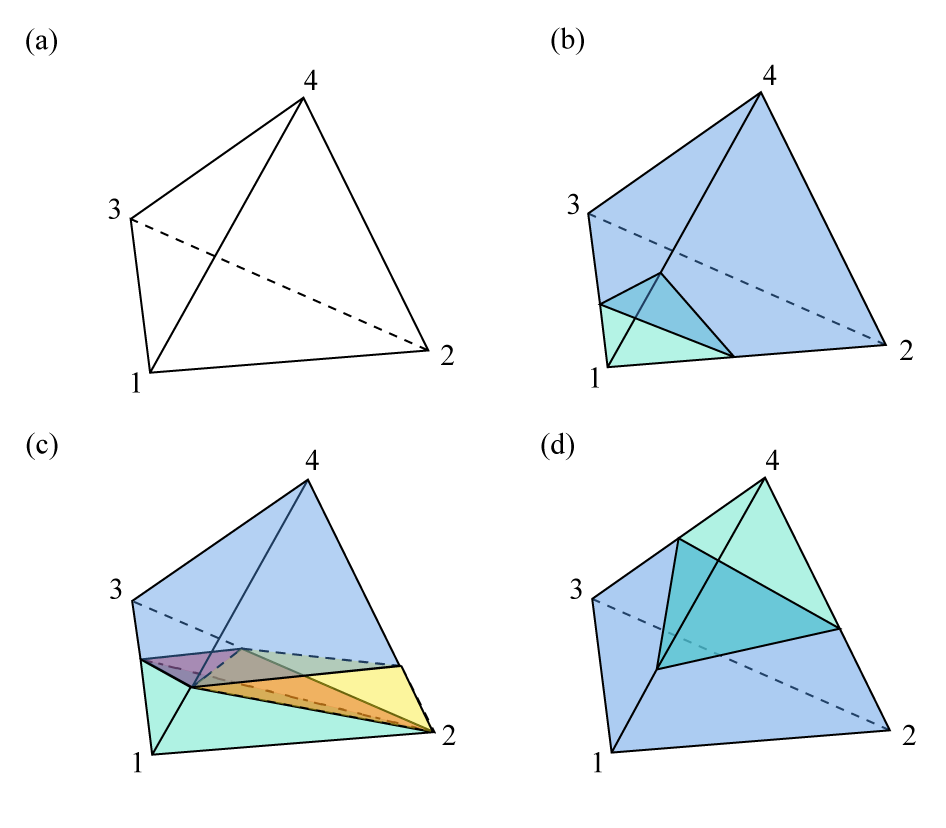}
    \caption{How to split tetrahedron. (a) A linear tetrahedron. Its vertices are labeled according to order of band eigenvalues, and $\varepsilon _{1} \leqslant \varepsilon _{2} \leqslant \varepsilon _{3} \leqslant \varepsilon _{4}$. (b) Case when $\varepsilon _{1} < \varepsilon _{F} \leqslant \varepsilon _{2}$, (c) Case when $\varepsilon _{2} < \varepsilon _{F} \leqslant \varepsilon _{3} $, (d) Case when $\varepsilon _{3} < \varepsilon _{F} \leqslant \varepsilon _{4} $ }
    \label{fig:split}
\end{figure}

Details of sub-tetrahedra are given in Table.\ref{tab:split}, including number of sub-tetrahedra, values at vertices and their volumes. We denote values at vertices of the original tetrahedron by $F_{1-4}$, and $F_{ij}$ is the linear-interpolated value at the crossing point of the Fermi surface with edge $\vec k_{i}-\vec k_{j}$, and is given by:
\[F_{ij}=\frac{\varepsilon _{j} -\varepsilon _{F}}{\varepsilon _{j} -\varepsilon _{i}} F_{i} +\frac{\varepsilon _{F} -\varepsilon _{i}}{\varepsilon _{j} -\varepsilon _{i}} F_{j}.\]

\begin{table}[!h]
    \centering
\caption{How to split a tetrahedron. Details of sub-tetrahedra in each case.\label{tab:split}}    
\begin{tabular}{|c|c|c|c|}
\hline 
case & number of sub-tetrahedra & values at vertices  & volume \\
\hline 
$\displaystyle \varepsilon _{1} < \varepsilon _{F} \leqslant \varepsilon _{2}$ & 1 & $\displaystyle \mathbf{f}_{1-4} =( F_{1} ,F_{12} ,F_{13} ,F_{14})$ & $\displaystyle V_{T}\frac{( \varepsilon _{F} -\varepsilon _{1})^{3}}{( \varepsilon _{2} -\varepsilon _{1})( \varepsilon _{3} -\varepsilon _{1})( \varepsilon _{4} -\varepsilon _{1})}$ \\
\hline 
&  & $\displaystyle \mathbf{f}^{1}_{1-4} =( F_{1} ,F_{2} ,F_{13} ,F_{14})$ & $\displaystyle V_{T}\frac{( \varepsilon _{F} -\varepsilon _{1})^{2}}{( \varepsilon _{4} -\varepsilon _{1})( \varepsilon _{3} -\varepsilon _{1})}$ \\
\cline{3-4} 
$\displaystyle \varepsilon _{2} < \varepsilon _{F} \leqslant \varepsilon _{3}$ & 3 & $\displaystyle \mathbf{f}^{2}_{1-4} =( F_{1} ,F_{13} ,F_{14} ,F_{23})$ & $\displaystyle V_{T}\frac{( \varepsilon _{F} -\varepsilon _{1})( \varepsilon _{F} -\varepsilon _{2})( \varepsilon _{3} -\varepsilon _{F})}{( \varepsilon _{4} -\varepsilon _{1})( \varepsilon _{3} -\varepsilon _{2})( \varepsilon _{3} -\varepsilon _{1})}$ \\
\cline{3-4} 
&  & $\displaystyle \mathbf{f}^{3}_{1-4} =( F_{1} ,F_{14} ,F_{23} ,F_{24})$ & $\displaystyle V_{T}\frac{( \varepsilon _{F} -\varepsilon _{2})^{2}( \varepsilon _{4} -\varepsilon _{F})}{( \varepsilon _{4} -\varepsilon _{2})( \varepsilon _{3} -\varepsilon _{2})( \varepsilon _{4} -\varepsilon _{1})}$ \\
\hline 
$\displaystyle \varepsilon _{3} < \varepsilon _{F} \leqslant \varepsilon _{4}$ & 1 & $\displaystyle \mathbf{f}_{1-4} =( F_{14} ,F_{24} ,F_{34} ,F_{4})$ & $\displaystyle V_{T}\frac{( \varepsilon _{4} -\varepsilon _{F})^{3}}{( \varepsilon _{4} -\varepsilon _{1})( \varepsilon _{4} -\varepsilon _{2})( \varepsilon _{4} -\varepsilon _{3})}$ \\
\hline
\end{tabular}
\end{table}

Consider a weight function of form $W(\vec k)=\Theta ( \varepsilon _{F} -\varepsilon (\mathbf{k})) \bar{W}(\vec k)$, we can apply the linear tetrahedron method of $\bar{W}(\vec k)$ to each sub-tetrahedron to express the its contribution as a weighted sum of its vertex values $f_{1-4}$ and further of $F_{1-4}$. In the case $\varepsilon _{3} < \varepsilon _{F} \leqslant \varepsilon _{4} $, of course, we have to subtract out the sub-tetrahedron contribution from the total contribution of the parent tetrahedron. As an example, the readers can check the linear tetrahedron rules for Eq.(\ref{theta}) listed below, where $\bar{W}(\vec k)=1$ and the contribution from each sub-tetrahedron is just the sum of its vertex values $f_{1-4}$ multiplied by a fourth of its volume.

\subsection{Linear tetrahedron rule for \texorpdfstring{$W(\vec k)=\Theta ( \varepsilon _{F} -\varepsilon (\mathbf{k}))$ and $\delta ( \varepsilon _{F} -\varepsilon (\mathbf{k}))$}{Linear tetrahedron rule for W(k) = Θ(εF−ε(k)) and δ(εF−ε(k))} \label{Wtheta}}

We consider the following integral within a linear tetrahedron $T$:
\[I=\int\limits_{T} d^d k \Theta ( \varepsilon _{F} -\varepsilon (\mathbf{k})) F(\mathbf{k}) \approx \sum _{i=1}^{4} \omega _{i} F_{i}.\]

There are five different situations depending on the Fermi energy $\varepsilon_F$:

\begin{enumerate}[(1)]
    \item For $\varepsilon _{F} \leqslant \varepsilon _{1}$ : 
    \[\omega _{1} =\omega _{2} =\omega _{3} =\omega _{4} =0\]
    \item For $\varepsilon _{1} < \varepsilon _{F} \leqslant \varepsilon _{2}$ :
\[\begin{aligned}
    \omega _{1} &=C\left[ 4-( \varepsilon _{F} -\varepsilon _{1})\left(\frac{1}{\varepsilon _{2} -\varepsilon _{1}} +\frac{1}{\varepsilon _{3} -\varepsilon _{1}} +\frac{1}{\varepsilon _{4} -\varepsilon _{1}}\right)\right]\\
    \omega _{2} &=C\frac{\varepsilon _{F} -\varepsilon _{1}}{\varepsilon _{2} -\varepsilon _{1}} ,\ \ \ \omega _{3} =C\frac{\varepsilon _{F} -\varepsilon _{1}}{\varepsilon _{3} -\varepsilon _{1}} ,\ \ \ \omega _{4} =C\frac{\varepsilon _{F} -\varepsilon _{1}}{\varepsilon _{4} -\varepsilon _{1}}
\end{aligned}\]
where 
    \[C\ =\ \frac{V_{T}}{4}\frac{( \varepsilon _{F} -\varepsilon _{1})^{3}}{( \varepsilon _{2} -\varepsilon _{1})( \varepsilon _{3} -\varepsilon _{1})( \varepsilon _{4} -\varepsilon _{1})}\]
    \item For $\varepsilon _{2} < \varepsilon _{F} \leqslant \varepsilon _{3} $:
    \[
\begin{aligned}
    \omega _{1} &=C_{1} +( C_{1} +C_{2})\frac{\varepsilon _{3} -\varepsilon _{F}}{\varepsilon _{3} -\varepsilon _{1}} +( C_{1} +C_{2} +C_{3})\frac{\varepsilon _{4} -\varepsilon _{F}}{\varepsilon _{4} -\varepsilon _{1}}\\  
    \omega _{2} &=C_{1} +C_{2} +C_{3} +( C_{2} +C_{3})\frac{\varepsilon _{3} -\varepsilon _{F}}{\varepsilon _{3} -\varepsilon _{2}} +C_{3}\frac{\varepsilon _{4} -\varepsilon _{F}}{\varepsilon _{4} -\varepsilon _{2}}\\
    \omega _{3} &=( C_{1} +C_{2})\frac{\varepsilon _{F} -\varepsilon _{1}}{\varepsilon _{3} -\varepsilon _{1}} +( C_{2} +C_{3})\frac{\varepsilon _{F} -\varepsilon _{2}}{\varepsilon _{3} -\varepsilon _{2}}\\
    \omega _{4} &=( C_{1} +C_{2} +C_{3})\frac{\varepsilon _{F} -\varepsilon _{1}}{\varepsilon _{4} -\varepsilon _{1}} +C_{3}\frac{\varepsilon _{F} -\varepsilon _{2}}{\varepsilon _{4} -\varepsilon _{2}}
\end{aligned}
    \]
    where 
    \[\begin{aligned}
        C_{1} &=\frac{V_{T}}{4}\frac{( \varepsilon _{F} -\varepsilon _{1})^{2}}{( \varepsilon _{4} -\varepsilon _{1})( \varepsilon _{3} -\varepsilon _{1})}\\
        C_{2} &=\frac{V_{T}}{4}\frac{( \varepsilon _{F} -\varepsilon _{1})( \varepsilon _{F} -\varepsilon _{2})( \varepsilon _{3} -\varepsilon _{F})}{( \varepsilon _{4} -\varepsilon _{1})( \varepsilon _{3} -\varepsilon _{2})( \varepsilon _{3} -\varepsilon _{1})}\\
        C_{3} &=\frac{V_{T}}{4}\frac{( \varepsilon _{F} -\varepsilon _{2})^{2}( \varepsilon _{4} -\varepsilon _{F})}{( \varepsilon _{4} -\varepsilon _{2})( \varepsilon _{3} -\varepsilon _{2})( \varepsilon _{4} -\varepsilon _{1})}
\end{aligned}\]
    \item For $\varepsilon _{3} < \varepsilon _{F} \leqslant \varepsilon _{4} $:
    \[\begin{aligned}
    \omega _{1} &=\frac{V_{T}}{4} -C\frac{\varepsilon _{4} -\varepsilon _{F}}{\varepsilon _{4} -\varepsilon _{1}}\\
    \omega _{2} &=\frac{V_{T}}{4} -C\frac{\varepsilon _{4} -\varepsilon _{F}}{\varepsilon _{4} -\varepsilon _{2}}\\
    \omega _{3} &=\frac{V_{T}}{4} -C\frac{\varepsilon _{4} -\varepsilon _{F}}{\varepsilon _{4} -\varepsilon _{3}}\\
    \omega _{4} &=\frac{V_{T}}{4} -C\left[ 4-( \varepsilon _{4} -\varepsilon _{F})\left(\frac{1}{\varepsilon _{4} -\varepsilon _{1}} +\frac{1}{\varepsilon _{4} -\varepsilon _{2}} +\frac{1}{\varepsilon _{4} -\varepsilon _{3}}\right)\right]
    \end{aligned}\]
    where 
    \[C\ =\ \frac{V_{T}}{4}\frac{( \varepsilon _{4} -\varepsilon _{F})^{3}}{( \varepsilon _{4} -\varepsilon _{1})( \varepsilon _{4} -\varepsilon _{2})( \varepsilon _{4} -\varepsilon _{3})}\]
    \item For $\varepsilon _{4} < \varepsilon _{F} $:
    \[\omega _{1} =\omega _{2} =\omega _{3} =\omega _{4} =\frac{V_{T}}{4}\]
\end{enumerate}

Note $C$ and $C_{1-3}$ are volumes of sub-tetrahedra in each case. In cases (2) and (3) we add contributions from sub-tetrahedra, while in case (4) we subtract out the sub-tetrahedron contribution from that of the parent tetrahedron $T$. The
linear tetrahedron rules for $W(\vec k)=\delta ( \varepsilon _{F} -\varepsilon (\mathbf{k}))$ can be obtained by $\omega_{1-4}$ above differentiated against $\varepsilon_F$ in each case, which is straightforward but lengthy and not given here.

\subsection{Linear tetrahedron rule for \texorpdfstring{$W(\vec k)=1/D(\mathbf{k})$}{Linear tetrahedron rule for W(k) = 1/D(k)}}
We consider the following integral within a linear tetrahedron $T$
\[\begin{aligned}
    & \int\limits_{T} d\mathbf{k} F(\mathbf{k}) /D(\mathbf{k}) \\
    \approx & 6V_{T}\int _{0}^{1} dx\int _{0}^{1-x} dy\int _{0}^{1-x-y} dz\ \frac{F_{1} +( F_{2} -F_{1}) x+( F_{3} -F_{1}) y+( F_{4} -F_{1}) z}{D_{1} +( D_{2} -D_{1}) x+( D_{3} -D_{1}) y+( D_{4} -D_{1}) z} \\
    =&\sum _{i=1}^{4} \omega _{i} F_{i}.
    \end{aligned}\]

$\omega_{1-4}$ is given by
\[\begin{aligned}
    \omega _{1} &=\phi ( D_{1} ,D_{2} ,D_{3} ,D_{4}) +\phi ( D_{1} ,D_{3} ,D_{4} ,D_{2}) +\phi ( D_{1} ,D_{4} ,D_{2} ,D_{3})\\
    \omega _{2} &=\phi ( D_{2} ,D_{3} ,D_{4} ,D_{1}) +\phi ( D_{2} ,D_{4} ,D_{1} ,D_{3}) +\phi ( D_{2} ,D_{1} ,D_{3} ,D_{4})\\
    \omega _{3} &=\phi ( D_{3} ,D_{4} ,D_{1} ,D_{2}) +\phi ( D_{3} ,D_{1} ,D_{2} ,D_{4}) +\phi ( D_{3} ,D_{2} ,D_{4} ,D_{1})\\
    \omega _{4} &=\phi ( D_{4} ,D_{1} ,D_{2} ,D_{3}) +\phi ( D_{4} ,D_{2} ,D_{3} ,D_{1}) +\phi ( D_{4} ,D_{3} ,D_{1} ,D_{2})
\end{aligned}\]
where 
\[\phi (a,b,c,d)=\frac{-\frac{a^{3}}{9} +\frac{a^{2} b}{4} -\frac{5b^{3}}{36} +\frac{1}{3} a^{3}\log |a|-\frac{1}{2} a^{2} b\log |a|+\frac{1}{6} b^{3}\log |b|}{(a-b)^{2} (b-c)(b-d)}.\]
$\phi (a,b,c,d)$ becomes singular when any of following equalities, $a=b$, $b=c$, $b=d$, is true. But in any case, $\omega_{1}$ has a proper limit when all $D_{1-4}$ are non-zeros. We list all limiting cases of $\phi (a,b,c,d)$ below:
\begin{enumerate}
    \item $a\rightarrow b$, $b\neq c \text{ or } d$
    \[\phi (a,b,c,d)\rightarrow\frac{b\log |b|}{2( b-c)( b-d)}\]
    \item $b\rightarrow c$, $c\neq a \text{ or } d$
    \[\begin{aligned}
        &\phi (a,b,c,d)+\phi (a,c,b,d)\\
        =&\frac{1}{36( a-c)^{3}( c-d)^{2}}\Bigl[( a-c)\Big( 4a^{3} +10ac( c-d) +c^{2}( 6c-d) -a^{2}( 8c+d)\Big) \\
        &-6a^{2}\left( 2a^{2} -6ac+6c^{2} +ad-3cd\right)\log |a|+6c^{2}( 2ac-3ad+cd)\log |c|\Bigr]
        \end{aligned}\]
    \item $a\&b\rightarrow c$, $a \neq d$
    \[\phi (a,b,c,d)+\phi (a,c,b,d)= -\frac{-c+d+( 2c+d)\log |c|}{6( c-d)^{2}}\]
    \item $b\&c\rightarrow d$, $a \neq d$
    \[\begin{aligned}
        &\phi (a,b,c,d)+\phi (a,c,d,b)+\phi (a,d,b,c)\\
        =&\frac{2a^{3} +3a^{2} d-6ad^{2} +d^{3} -6a^{2} d\log |a|+6a^{2} d\log |d|}{12( a-d)^{4}}
    \end{aligned}\]
    \item $a\&b\&c\rightarrow d$
    \[\phi (a,b,c,d)+\phi (a,c,d,b)+\phi (a,d,b,c)=\frac{1}{24d}\] 
\end{enumerate}

Case $1$ can come simultaneously with case $2$, for example, when $a\rightarrow d$ and $b\rightarrow c$ which means both $\phi(a,d,b,c)$ and $\phi(a,b,c,d)+\phi(a,c,b,d)$ have limits to take care of. The calculation of integral weights is susceptible to precision loss close to any of these limiting cases. If any pair of denominators are very close to each other, the weights are better calculated assuming they are equal and use the limits above. The problem comes when denominators are close, but either the limits at equalites or ordinary calculation is satisfactorily accurate, and we suggest calculating ordinarily with a precision promotion to quad precision floating-point number in this situation.

For possible applications in 2D, we also present the results for "the linear triangle method". We consider the following integral within a triangle $T$
\[\begin{aligned}
    &\int\limits_{T} d\mathbf{k} F(\mathbf{k}) /D(\mathbf{k}) \\
    \approx &6V_{T}\int _{0}^{1} dx\int _{0}^{1-x} dy\ \frac{F_{1} +( F_{2} -F_{1}) x+( F_{3} -F_{1}) y}{D_{1} +( D_{2} -D_{1}) x+( D_{3} -D_{1}) y} =\sum _{i=1}^{3} \omega _{i} F_{i}
\end{aligned}\]
$\omega_{1-3}$ is given by
\[\begin{aligned}
    \omega _{1} &=\phi ( D_{1} ,D_{2} ,D_{3}) +\phi ( D_{1} ,D_{3} ,D_{2})   \\
    \omega _{2} &=\phi ( D_{2} ,D_{3} ,D_{1}) +\phi ( D_{2} ,D_{1} ,D_{3})   \\
    \omega _{3} &=\phi ( D_{3} ,D_{1} ,D_{2}) +\phi ( D_{3} ,D_{2} ,D_{1})
\end{aligned}\]
where
\[\phi (a,b,c)=-\frac{b}{2(a-b)(c-b)} -\frac{b^{2} (-\log |a|+\log |b|)}{2(a-b)^{2} (c-b)}\]

Similar to the 3D case, $\phi (a,b,c)$ becomes singular when $a=b$ or $b=c$, but $\omega_{1}$ has a proper limit unless any of $D_{1-3}$ is zero. We list below all limiting cases for $\phi (a,b,c)$ below:

\begin{enumerate}
    \item $a\rightarrow b$, $b\neq c$
    \[\phi (a,b,c)=\frac{1}{4b-4c}\]
    \item $b\rightarrow c$, $a\neq c$
    \[\phi (a,b,c)+\phi (a,c,b)=\frac{a^{2} -c^{2} -2ac\log |a|+2ac\log |c|}{2( a-c)^{2}}\]
    \item $a\&b\rightarrow c$
    \[\phi (a,b,c)+\phi (a,c,b)=\frac{1}{6c}\]
\end{enumerate}

As in the 3D case, we should take care of the situation when any pair of denominators are close to each other.
If $D(\vec{k})$ is complex, the above formulas still hold after changing the arguments of logarithms from absolute values of $D(\vec{k})$ to complex numbers $D(\vec{k})$ itself.    
  
  \end{widetext}

%

\end{document}